\documentclass[12pt,preprint]{aastex}
\usepackage{graphicx}

\slugcomment{--}

\shorttitle{Multiple stellar populations in 47 Tucanae} 
\shortauthors{Milone et al.} 

\begin{document}

\title{ Multiple Stellar Populations in 47 Tucanae.
          \footnote{           Based on observations with  the
                               NASA/ESA {\it Hubble Space Telescope},
                               obtained at  the Space Telescope Science
                               Institute,  which is operated by AURA, Inc.,
                               under NASA contract NAS 5-26555.}}

\author{
A.\ P. \,Milone\altaffilmark{2,3},
G.\ Piotto\altaffilmark{4},
L.\ R. \,Bedin\altaffilmark{5},
I.\ R. \,King\altaffilmark{6}, 
J.\ Anderson\altaffilmark{5}, 
A.\ F. \,Marino\altaffilmark{7}, 
A.\ Bellini\altaffilmark{4},
R.\ Gratton\altaffilmark{8},
A.\ Renzini\altaffilmark{8},
P.\ B. \,Stetson\altaffilmark{9},
S.\ Cassisi\altaffilmark{10}, 
A.\ Aparicio\altaffilmark{2,3}, 
A.\ Bragaglia\altaffilmark{11},
E.\ Carretta\altaffilmark{11},
F.\ D'Antona\altaffilmark{12},
M.\ Di Criscienzo\altaffilmark{12},
S.\ Lucatello\altaffilmark{8},
M.\ Monelli\altaffilmark{2,3}, and
A.\ Pietrinferni\altaffilmark{10}
 }

\altaffiltext{2}{Instituto de Astrof\`\i sica de Canarias, E-38200 La
              Laguna, Tenerife, Canary Islands, Spain; [milone,
                aparicio, monelli]@iac.es}

\altaffiltext{3}{Department of Astrophysics, University of La Laguna,
           E-38200 La Laguna, Tenerife, Canary Islands, Spain}

\altaffiltext{4}{Dipartimento  di   Astronomia,  Universit\`a  di Padova,
           Vicolo dell'Osservatorio 3, Padova I-35122, Italy;
           [giampaolo.piotto, andrea.bellini]@unipd.it }

\altaffiltext{5}{Space Telescope Science Institute,
                  3800 San Martin Drive, Baltimore,
                  MD 21218; [jayander,bedin]@stsci.edu}

\altaffiltext{6}{Department of Astronomy, University of Washington, Box
           351580, Seattle, WA 98195-1580; king@astro.washington.edu }

\altaffiltext{7}{Max Plank Institute for Astrophysics, Postfach 1317,
	   D-85741 Garching, Germany; amarino@MPA-Garching.MPG.DE}

\altaffiltext{8}{INAF-Osservatorio Astronomico di Padova, Vicolo
             dell'Osservatorio 5, I-35122 Padua, Italy;
             [raffaele.gratton, alvio.renzini, sara.lucatello]@oapd.inaf.it}

\altaffiltext{9}{Dominion Astrophysical Observatory, Herzberg Institute of
	   Astrophysics, National Research Council, 5071 West Saanich
	   Road, Victoria, British Columbia V9E 2E7, Canada; 
           Peter.Stetson@nrc-cnrc.gc.ca}

\altaffiltext{10}{INAF-Osservatorio Astronomico di Collurania, via Mentore
           Maggini, I-64100 Teramo, Italy [cassisi,
             pietrinferni]@oa-teramo.inaf.it} 

\altaffiltext{11}{INAF, Osservatorio Astronomico di Bologna, via Ranzani
           1, I-40127 Bologna, Italy; 
           [angela.bragaglia, eugenio.carretta]@oabo.inaf.it}

\altaffiltext{12}{INAF-Osservatorio Astronomico di Roma, Via Frascati 33,
	   I-00040 Monte Porzio Catone, Rome, Italy;
           dantona@mporzio.astro.it, dicrisci@gmail.com}

\begin{abstract}
 We use {\it Hubble Space Telescope (HST)} and ground-based imaging to
   study the multiple populations of 47 Tuc, combining high-precision
   photometry with calculations of synthetic spectra.  Using filters
   covering a wide range of wavelengths, our {\it HST} photometry splits
   the main sequence into two branches, and we find that this duality is
   repeated in the subgiant and red-giant regions, and on the horizontal
   branch.  We calculate theoretical stellar atmospheres for main
   sequence stars, assuming different chemical composition mixtures, and
   we compare their predicted colors through the {\it HST} filters with
   our observed colors.  We find that we can match the complex of observed
   colors with a pair of populations, one with primeval abundance and
   another with enhanced nitrogen and a small helium enhancement, but
   with depleted C and O.  We confirm that models of red giant and red
   horizontal branch stars with that pair of compositions also give
   colors that fit our observations. We suggest that the different
   strengths of molecular bands of OH, CN, CH and NH, falling in
   different photometric bands, are responsible for the color splits of
   the two populations. Near the cluster center, in each portion of the
   color-magnitude diagram the population with primeval abundances makes
   up only $\sim 20\%$ of the stars, a fraction that increases outwards,
   approachng equality in the outskirts of the cluster, with a fraction
   $\sim 30\%$ averaged over the whole cluster. Thus the second,
   He/N-enriched population is more concentrated and contributes the
   majority of the present-day stellar content of the cluster.  We
   present evidence that the color-magnitude diagram of 47 Tuc consists
   of intertwined sequences of the two populations, whose separate
   identities can be followed continuously from the main sequence up to
   the red giant branch, and thence to the horizontal branch.  A third
   population is visible only in the subgiant branch, where it includes
   $\sim 8\%$ of the stars.
\end{abstract}


%
\section{Introduction}
\label{introduction}
The presence of multiple stellar populations in globular clusters (GCs)
has been widely established both by photometric and by spectroscopic
studies.  More than 40 years ago the red giant branch (RGB) of $\omega$ 
Centauri (NGC 5139) was found to have a photometric spread in color (Woolley
1966), associated with a metallicity spread (Freeman \& Rodgers 1975, 
Norris \& Bessel 1975),
but the first real challenge to the traditional picture of
GCs as simple stellar populations came from the discovery of
abundance anomalies among stars within the same GCs (e.g., Kraft 1979).
All Galactic GCs studied so far show Na-O anti-correlations (e.g.,
Carretta et al.\ 2009a), indicative of contamination from products of
proton-capture reactions at high temperature (Denisenkov \& Denisenkova
1989, Langer et al.\ 1993).  Moreover, the presence of such
anti-correlations in unevolved main sequence (MS) stars (e.g., Gratton
et al.\ 2001; Ramirez \& Cohen 2002), which have not yet reached
sufficiently high temperatures in their interiors, suggests that more
than one generation of stars has formed within Galactic GCs (see 
Gratton et al.\ 2004 for a review), possibly in combination with
accretion onto low-mass stars of ejecta from either intermediate-mass
stars (D'Antona et al.\ 1983; Renzini 1983), fast-rotating massive stars
(e.\ g.\ , Decressin et al.\ 2007), or massive binaries (De Mink et
al.\ 2009).

Clear evidence of a complex star-formation history in GCs has come from
high-precision Hubble Space Telescope ({\it HST}) photometry, which
showed unequivocally that the phenomenon of multiple sequences in the
color-magnitude diagrams (CMDs) of GCs is not confined to the special
case of $\omega$ Centauri (for which see, e.g., Anderson 1997, Bedin et
al.\ 2004).  Split or spread MSs have been observed in NGC 2808, 47
Tucanae (NGC 104), and NGC 6752 (Piotto et al.\ 2007, Anderson et
al.\ 2009, Milone et al.\ 2010), while splits in the subgiant branches
(SGBs) have been detected in NGC 1851, NGC 6656 (M22), 47 Tuc, and at
least five other GCs (Milone et al.\ 2008, Marino et al.\ 2009, Anderson
et al.\ 2009, Piotto 2009).  The splitting of the red-giant branch (RGB)
has been observed in all the GCs studied to date with appropriate
photometric bands (e.g., Marino et al.\ 2008; Yong et al.\ 2008, Lee at
al.\ 2009, Lardo et al.\ 2011).  The photometric investigations
definitively confirm that it is common for GCs to contain multiple
stellar populations.

More recently it has been possible to connect the photometrically
observed multiple sequences with differences of chemical composition
among the stars in the same cluster.  The aim of such studies has been
to understand how successive generations of stars could have formed in a
GC, and what could be inferred about the nature of the polluters from
their chemical imprint on the second-generation stars.  The first
notable results were by Piotto et al.\ (2005), who showed that the bluer
MS in $\omega$ Centauri is more metal-rich than the redder one.  The
only way to reconcile the photometric and spectroscopic results is to
assume that the bluer MS is strongly He-enhanced. Other examples are
given by
Marino et al.\ (2008), who showed that the presence of two groups of
stars with different C, N, Na, and O content is at the root of the
difference in the $(U-B)$ color of the RGB stars in NGC 6121 (M4).
Na-poor (CN-weak) giants define a sequence bluer than the one occupied
by the Na-rich (CN-strong) ones.  Yong et al.\ (2008) found that RGB
spreads are present in a large number of GCs, when using the
Str\"{o}mgren $c1$ index, which is a powerful tracer of the N abundance.
In general, stellar evolutionary models
suggest that observed multiple MSs should correspond to stellar
populations with different He abundance, the bluer sequences having
a higher helium abundance than primordial (e.g., Norris
2004; D'Antona et al.\ 2005; Piotto et al.\ 2005, 2007; Di Criscienzo et
al.\ 2010, 2011).
According to this picture, the triple MS in NGC 2808 can also correspond
to the three groups of stars with different oxygen and sodium abundances
observed among the RGB stars (Carretta et al.\ 2006), likely to come
from three successive episodes of star formation.  In fact, hydrogen
burning at high temperatures through the CNO cycle and subsequent proton
captures result in a He enrichment, and at the same time in an
enhancement of N, Na, and Al, and a depletion of C, O, and Mg.
This scenario has been nicely confirmed for NGC 2808 by Bragaglia et
al.\ (2010), who measured chemical abundances of
one star on the red MS and one on the blue MS,
and found that the latter shows an enhancement of N, Na, and Al, and a
depletion in C and Mg.
To date there are no spectroscopic studies of stars on the middle branch
of the MS, however.

Finally, theoretical models have proposed that the SGB split observed in
some GCs could be due to two stellar groups with
either an age difference of 1--2 Gyr or a different C+N+O content
(Cassisi et al.\ 2008, Ventura et al.\ 2009).  In support of the
chemical-content scenario, a bimodality in the $s$-process elements and
in the CNO has been detected in NGC 1851 and M22 (Yong et al.\ 2009,
Marino et al.\ 2011a).

Summarizing the observational and theoretical scenarios:
\begin{itemize}
\item{split MSs suggest helium enrichment, and also
correspond to different groups of stars in the Na-O anticorrelation;}
\item{
multiple SGBs might be ascribed to different total abundance of C+N+O or
to different age;}
\item{photometrically multiple RGBs might be due to differences in C, N,
O content, via the different strengths of the corresponding molecular
features;}
\item{there appears to be no strong indication of any significant spread
  in [Fe/H] (except for a few clusters:\ $\omega$ Centauri, M22,
  Terzan 5, and NGC 2419).}
\end{itemize}

Multiple stellar populations with different helium abundance also offer
an explanation for the complex, extended, and clumpy HB morphology
exhibited by some clusters (e.g., D'Antona et al.\ 2005, D'Antona \&
Caloi 2008, Catelan, Valcarce, \& Sweigart 2009, Gratton et al.\ 2010).
A direct confirmation of a connection of the HB shape with the
chemical content of the HB comes from recent work by Marino et al.\
(2011b), who have found that stars on the blue side of the instability
strip of the cluster M4 are Na-rich and O-poor, whereas stars on the red
HB are all Na-poor.

While the presence of multiple stellar populations in GCs as revealed by
multiplicities of the MS, SGB, RGB, or HB has been clearly established
in many clusters, efforts to unequivocally connect the various
evolutionary stages of each individual stellar generation have met with
only modest success so far, because individual sequences appear to merge
and even cross each other in some parts of the CMD, strongly depending
on the photometric bands used to build the CMD.  Connecting the various
branches from the MS to the HB would greatly help to fully characterize
each individual stellar generation, in terms of composition and age.

In the present paper we attack this problem by applying high-precision
{\it HST} photometry to the globular cluster 47 Tucanae 
(GO-12311, PI Piotto).  
This is one of the GCs in the Milky Way where multiple stellar
populations have recently been detected and studied both photometrically
and spectroscopically.  From the analysis of a large number of archival
{\it HST} images of the inner $\sim 3\times3$ arcmin, Anderson et
al.\ (2009) found that the SGB is spread in magnitude, with at least two
distinct branches:\ a brighter one with an intrinsic broadening in
luminosity, and a second one about 0.05 mag fainter that 
includes a small fraction of the stars.
Anderson et al.\ were also able to study the MS in a 
            less crowded field 6 arcmin from the center and found
            an intrinsic broadening that increases towards fainter
            magnitudes. 
They interpreted the MS spread in terms of a variation of helium abundance of
$\sim 0.02$--0.03.  Di Criscienzo et al.\ (2010) suggested that a
spread in helium of $\sim 0.02$ might be responsible for both the
luminosity spread of the bright SGB and the HB morphology, whereas an
increase in the overall C+N+O abundance could be responsible for the
faint SGB.
In substantial agreement with the above estimates of the differences in
helium content in 47 Tuc, Nataf et al. (2011) have estimated a helium
difference of $\Delta Y\simeq 0.03$ between two sub-populations of this
cluster, based on the strength and luminosity of the RGB bump and on the
luminosity of the HB.  They also noted that the helium-rich population
is more centrally concentrated.

Since the early seventies, spectroscopic investigations have shown that
RGB stars in 47 Tuc exhibit large star-to-star variations in CN band
strength (e.g., McClure \& Osborn \ 1974, Bell, Dickens, \& Gustafsson
1975), with two distinct groups of stars showing different CN content
(Norris \& Freeman 1979, Briley 1997); this dichotomy is also present
among MS stars (Cannon 1998, Harbeck et al.\ 2003).
A Na--O anticorrelation has recently been studied by Carretta et
al.\ (2009a,b), with $\sim 30$\% of the stars being Na-poor and O-rich,
while the remaining $\sim 70$\% are depleted in oxygen and enhanced in
sodium.

This paper is organized as follows: In Section~\ref{data} we describe the
data and the reduction.  Section~\ref{sec:MS} reveals a split in the MS,
and in Section~\ref{YcnoMS} we explore possible theoretical
interpretations.  Sections~\ref{sec:SGB}, \ref{sec:RGB}, and
\ref{sec:HB} return to the pursuit of multiple sequences along 
the SGB, the RGB, and the HB respectively, and also explore their
interpretation.  The spatial distribution of multiple stellar
populations is investigated in Section~\ref{sec:RD}, while in
Section~\ref{sec:con} we attempt to connect the multiple sequences that
we have found along the MS, the SGB, the RGB, and the HB, and to trace
the CMD of each of the two stellar generations. A summary
and some final discussion follow in Section~\ref{sec:discussion}.

\section{Observations and data reduction}
\label{data}

For our study of the stellar populations in 47
Tuc, we used data sets from two different telescopes.  For the crowded
central regions of the cluster we used {\it HST} images taken with the
Wide Field Channel of the Advanced Camera for Surveys (ACS/WFC) and the
UVIS channel of Wide Field Camera 3 (WFC3/UVIS), while to study the
spatial distribution of the populations we made use 
of {\it U}, {\it B}, {\it V}, and {\it I} ground-based photometry from
the data base of 856 original and archival CCD images from Stetson
(2000).  Among them, 480 images were obtained with the Wide-Field Imager
of the ESO/MPI 2.2 m telescope, and 200 with the 1.5 m telescope at
Cerro Tololo Inter-American Observatory, while the remaining 176 images
come from various other telescopes.  These observations are described in
detail in Bergbusch \& Stetson (2009). They were reduced following the
protocol outlined in some detail by Stetson (2005) and are calibrated on
the Landolt (1992) photometric system.

\begin{figure}[ht!]
   \centering

   \epsscale{.75}
   \plotone{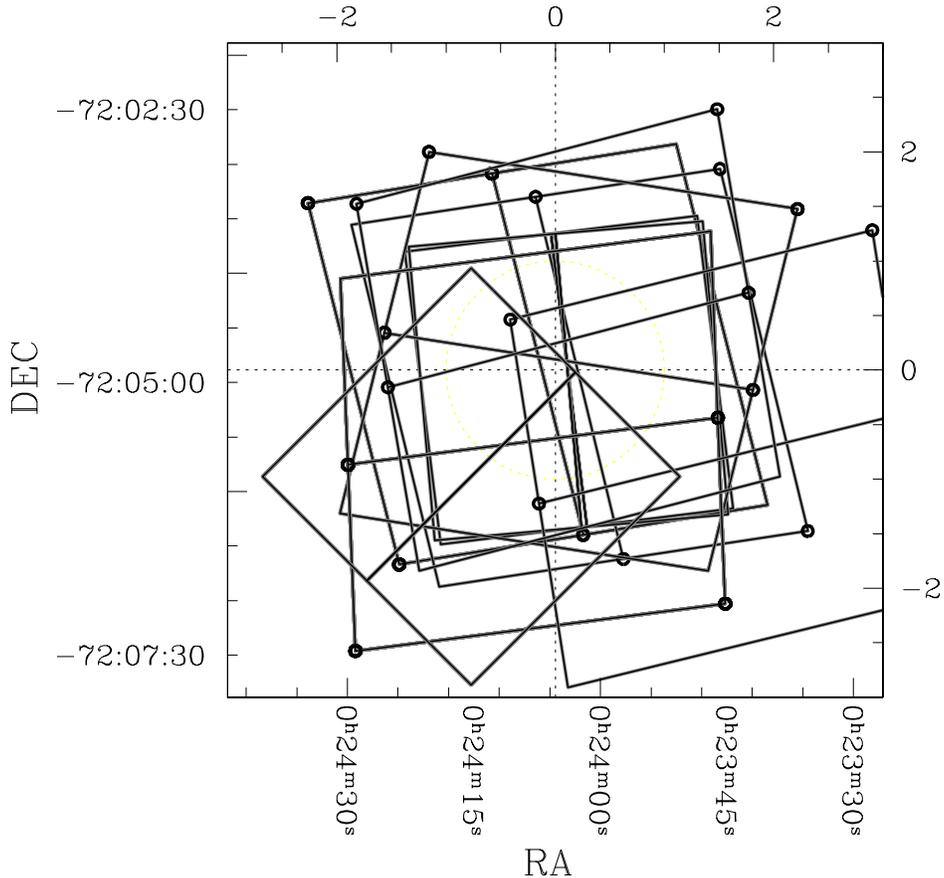}
\caption{Footprints of our {\it HST} fields.  The small circles are the
  corners of ACS/WFC chip 1.  The footprints of WFC3/UVIS exposures are
  distinguished by having no such marking.}
         \label{footprint}
   \end{figure}

Table~1 summarizes the characteristics of the {\it HST} images that we
used, while Fig.~\ref{footprint} shows their footprints.
The ACS images, and a number of the WFC3 images, were archival, except
that the images of GO-12311 (PI Piotto) were taken expressly for
this project, and were crucial to its success.

%
The ACS/WFC images were reduced by using the software described in
Anderson et al.\ (2008).  It consists of a package that analyzes all the
exposures simultaneously to generate a catalog of stars over the whole
field of view.  Stars are measured in each image independently by using
for each filter a spatially variable point-spread-function model from
Anderson \& King (2006) plus a ``perturbation PSF'' that allows for the
effects of focus variations.  The photometry was put into the Vega-mag
system following the recipes of Bedin et al.\ (2005) and using the
encircled energy and zero points of Sirianni et al.\ (2005).

Star positions and fluxes in the WFC3 images were measured with software
that is mostly based on img2xym\_WFI (Anderson et al.\ 2006); 
this will be
presented in a separate paper. Star positions and fluxes were corrected
for pixel area and geometric distortion by using the solution given by
Bellini \& Bedin (2009)
 and Bellini, Anderson \& Bedin (2011), and
were calibrated as in Bedin et al.\ (2005). 

The work that we present here is based mainly on high-precision
photometry, for which our next step was to select a high-quality sample
of stars that are relatively isolated and have small photometric and
astrometric errors, and  are also well fit by the PSF.
For this we used the quality indices that our photometry software
produces, in a procedure that is described in detail by Milone et
al.\ (2009, Sect.\ 2.1).  Finally we corrected our photometry for some
remaining position-dependent errors, due to small inadequacies in our
PSFs that were quite small but were different for each filter.  Since
all the uses of our photometry would depend on colors, we generated the
36 colors that can be derived from our 9 filters.  For each of these
colors we drew the main-sequence ridgeline in the corresponding CMD;
then for each star we identified its 50 closest well-measured neighbors,
and found their median color offset from the main-sequence ridgeline.
Since this constituted a good estimate of the systematic color error at
the position of the target star, for that pair of filters we corrected
the observed color of the star by that amount.

\begin{table*}[ht!]
\begin{center}  
\caption{{\it HST} data sets used in this paper. }
\scriptsize {
\begin{tabular}{lccccc}
\hline
\hline
 INSTR &  DATE & N$\times$EXPTIME & FILT  & PROGRAM &PI \\
\hline
 UVIS/WFC3 & Nov 21-22 2011       & 2$\times$323s+12$\times$348s & F275W   & 12311 & Piotto \\
 UVIS/WFC3 & Sep 28 2010          & 30s+1160s                    & F336W   & 11729 &Holtzman \\
 UVIS/WFC3 & Sep 28 2010          & 2$\times$10s$+$2$\times$348s+2$\times$940s & F390W & 11664 &Brown \\
 ACS/WFC   & Sep 30-Oct 11 2002   &   9$\times$105s              & F435W  &  9281 &Grindlay \\
 ACS/WFC   & Apr 05 2002          &  20$\times$60s               & F475W  &  9028 &Meurer \\
 ACS/WFC   & Jul 07 2002          &   5$\times$60s               & F475W  &  9443 &King  \\
 ACS/WFC   & Jul 07 2002          &   1$\times$150s              & F555W  &  9443 &King \\
 ACS/WFC   & Mar 13 2006          &  3s$+$4$\times$50s           & F606W  & 10775 &Sarajedini \\
 ACS/WFC   & Sep 30-Oct 11 2002   &  20$\times$65s               & F625W  &  9281 &Grindlay \\
 ACS/WFC   & Mar 13 2006          &  3s$+$4$\times$50s           & F814W  & 10775 &Sarajedini \\
\hline
\hline
\end{tabular}
}
\end{center}
\label{tabdata}
\end{table*}

\section{The double Main Sequence}
\label{sec:MS}

As already mentioned in Section~\ref{introduction}, 
the initial  evidence that the
MS of 47 Tuc is not consistent with a single stellar population comes
from the recent work by Anderson et al.\ (2009), who detected an
intrinsic spread in the $m_{\rm F606W}-m_{\rm F814W}$ color ranging from
$\sim 0.01$ mag near $m_{\rm F606W}\sim$ 19.0 to $\sim 0.02$ mag around
$m_{\rm F606W}=22.0$.  
Our present data set, however, allowed us to study the main sequence
with higher precision than was possible for Anderson et al. with the
data available at that time.

\begin{figure*}[ht!]
  \centering 
   \epsscale{.75}
   \plotone{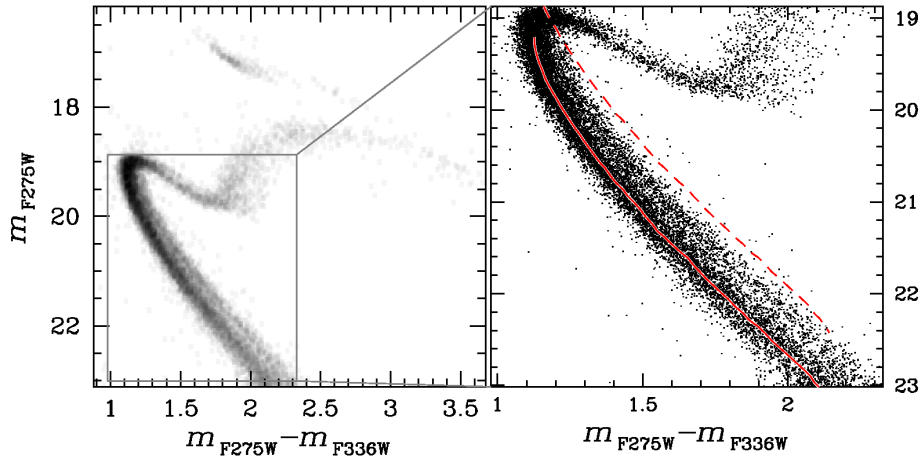}
     \caption {$m_{\rm F275W}$ vs.\ $m_{\rm F275W}-m_{\rm F336W}$ Hess
      diagram (\textit {left}), and CMD zoomed around the MS region
      (\textit {right}). The continuous and the dashed red lines in the
      right panel mark the MS ridge line of MSb and the equal-mass
      binary sequence, respectively.}
        \label{MS}
  \end{figure*}

An inspection of the large number of CMDs that we obtain from the data
set listed in Table 1 showed that the multiple populations along the MS
are best recognized and separated from photometry that combines ${\it
m}_{\rm F275W}$ with $m_{\rm F336W}$.
The left-hand panel of Fig.~\ref{MS} shows the Hess diagram of $m_{\rm
F275W}$ vs.\ $m_{\rm F275W}-m_{\rm F336W}$, after the quality selection
and photometric corrections described in the previous Section.  We
immediately note a widely spread RGB, a bimodal SGB, and a double MS.
To examine these more closely, in the right-hand half of the figure we
show a CMD that is zoomed around the upper MS, the SGB, and the start of
the RGB.  We defer discussion of the SGB and RGB to later Sections, and
concentrate here on the main-sequence morphology.

\begin{figure*}[ht!]
   \centering
   \epsscale{.75}
   \plotone{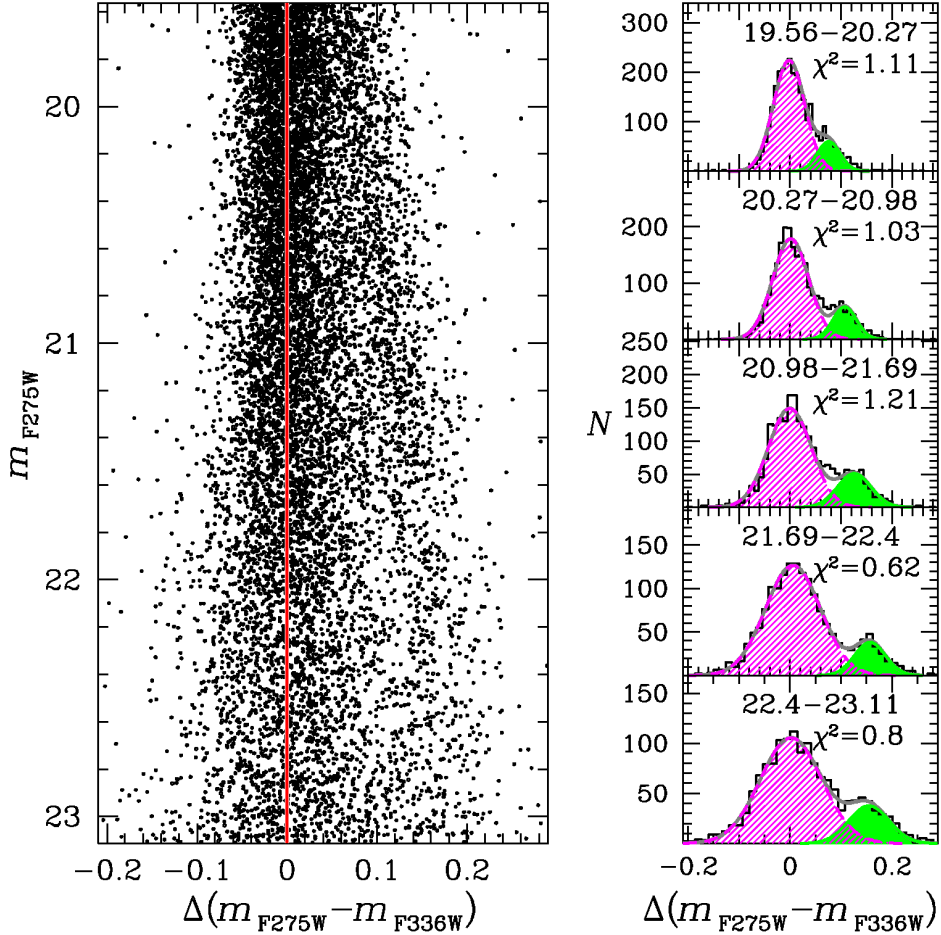}
      \caption{\textit {Left:} The same CMD as in Fig.~\ref{MS}, but
        after subtracting from the color of the MS the color of the
        ridge line of MSb.  \textit {Right:} Color distributions of the
        points in the left panel, showing two clear peaks. The magenta
        and green solid lines are the least-squares fits of two
        Gaussians to the histograms, while the gray line is their
        sum. 
        We have also indicated the reduced-$\chi^2$ value
        corresponding to each bi-Gaussian fit.
      }
         \label{MSnor}
   \end{figure*}

The CMD in the 
right-hand panel suggests that the MS of 47 Tuc is bimodal, in analogy
with the multiple MSs observed in $\omega$ Centauri, NGC 2808, and NGC
6752.  In 47 Tuc, however, the majority of MS stars populate the blue
component (hereafter MSb), while a small but significant fraction of the
stars lie on a redder MS branch (MSa).  The two sequences merge close to
the turn-off, but on the MS the separation of the two components
increases 
towards fainter magnitudes, from 0.08 mag at $m_{\rm F275W}=19.5$ to 0.15
mag at $m_{\rm F275W}=23$ --- as illustrated in more detail in
Figure~\ref{MSnor}.  Such
large, clear separations allow us to exclude any possibility that the MS
split might be due to measuring errors.

We note here that the ``verticalizing'' of the MS in Fig.~\ref{MSnor} is
a process that we will carry out several times in the course of
this paper, and we now explain it once and for all:  We first designated
MSb
as our target sequence, by means of a hand-drawn first approximation to
its ridge line, and we also chose a limited color range around this
line.  We then put a spline through the median colors in successive
short intervals of magnitude, and did an iterated sigma-clipping of
outliers; the result was a fiducial sequence for MSb.  The
verticalization then consisted of subtracting from the color of each
star the color of the fiducial sequence at the magnitude of that star.

Fig.~\ref{MSnor} allows us to estimate the fractions of MSa and MSb stars
by assigning to each star a verticalized color (left panel of the
figure), as explained above.  The right-hand panels of the figure
show the color histograms for five magnitude bins; in each bin we fitted
the histogram with a pair of Gaussians, shown in magenta and
green. These colors will be used consistently hereafter, to distinguish
these two sequences and their post-MS progeny.  From the areas under the
Gaussians we estimate that 82\% of the stars belong to MSb and 18\% to
MSa; within the statistical uncertainties of these numbers they have the
same values in each of the magnitude intervals.

   \begin{figure*}[ht!]
   \centering
   \epsscale{.75}
   \plotone{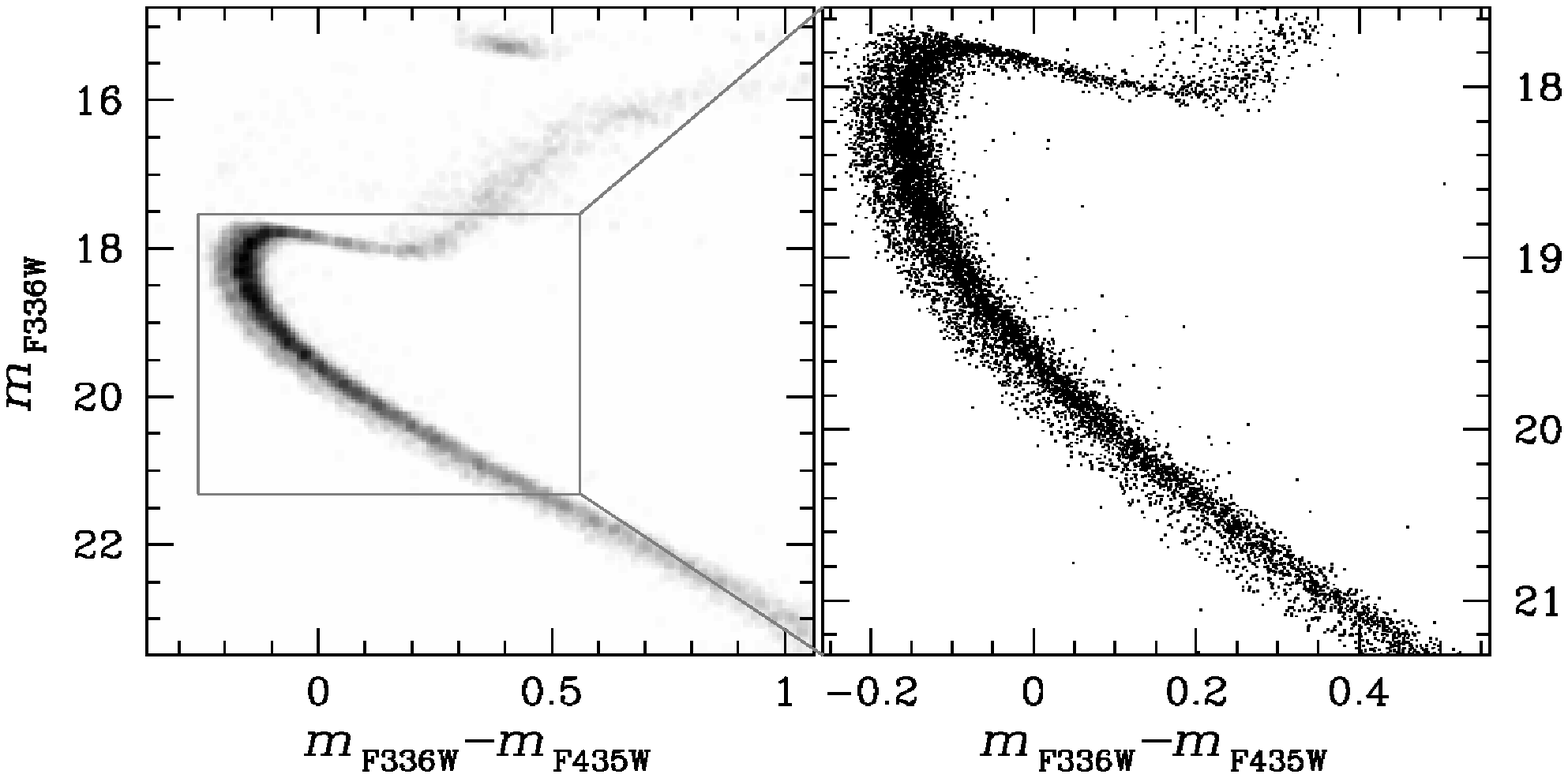}
      \caption{ $m_{\rm F336W}$ vs.\ $m_{\rm F336W}-m_{\rm F435W}$ Hess
       diagram (\textit {left panel}) and CMD zoomed around the region
       where the split is most evident (\textit {right panel}). Note
       that, contrary to its behavior in the CMD of Fig.~\ref{MS}, the
       less populous MS is bluer here than the bulk of MS stars.}
         \label{MSub}
   \end{figure*}

We also note that MSa cannot be ascribed to a sequence of binaries.  The
dashed line in Fig.~\ref{MS} is the equal-mass-binary sequence that
corresponds to the fiducial sequence of MSb.
Interpreting the stars of MSa as binaries would require making
the outlandish hypothesis that about a fourth of the MS stars in 47 Tuc
are in binary systems with mass ratio in the narrow interval 0.7--0.8.
 In addition, assuming that all MSa stars are binaries would imply a
binary fraction ($\ge$15\%), in sharp contrast with recent estimates of a
2\% binary fraction by Milone et al.\ (2008, 2011)\footnote{We note
  that the larger binary fraction for 47 Tuc proposed by Albrow et
  al.\ (2001) comes from extrapolation from the fraction of W UMa stars
  that they found in the same cluster, based on assumptions about the
  distribution of binary periods, and W UMa binary evolution}.  In view
of this, we expect that binaries do not affect the following discussion
in any significant way.

The double MS is 
also evident in CMDs that use a different combination of magnitude and
color, as shown in the $m_{\rm F336W}$ vs.\ $m_{\rm F336W}-m_{\rm
F435W}$ Hess diagram and CMD of Fig.~\ref{MSub}.  
It is important to note that in this color system the less populous MS
component is {\it bluer} than the other MS stars.

\begin{figure*}[ht!]
   \centering
   \epsscale{0.75}
   \plotone{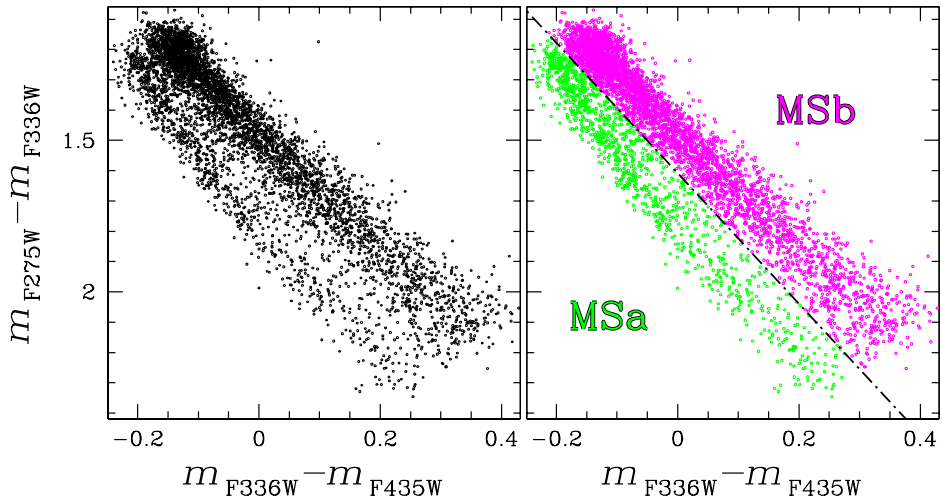}
      \caption{The $m_{\rm F275W}-m_{\rm F336W}$ vs.\ ${\it m}_{\rm
          F336W}-m_{\rm F435W}$ two-color diagram for MS stars with
        $19.56<m_{\rm F275W}<23.11$. The dot-dash line in the right-hand
        panel separates the MSa and MSb stars, shown in green and
        magenta colors, respectively.}
         \label{2C}
   \end{figure*}

Since these two CMDs from the filter set F275W, F336W, F435W behave so
differently, we construct the two-color diagram that is shown on the
left side of Figure~\ref{2C}.  There is a clear separation between the
two populations, and in the right panel the stars on opposite sides of
the dividing line have been colored green and magenta to identify
stars of MSa and MSb, respectively.
%

\begin{figure*}[ht!]
   \centering
   \epsscale{0.75}
   \plotone{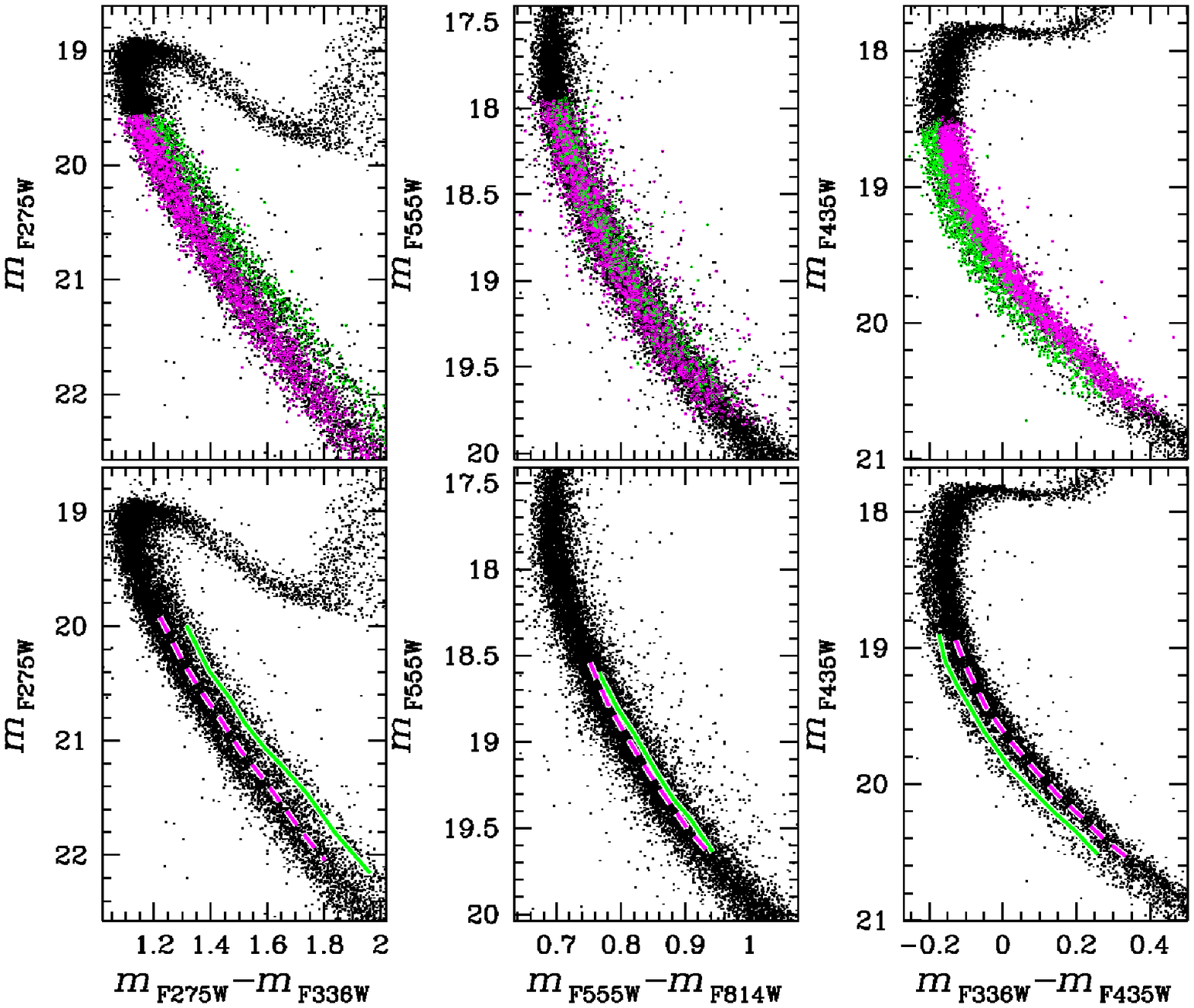}
      \caption{Example of the definition of MS fiducials. Upper panels
        show three different CMDs zoomed around the MS, with the MSa and
        MSb stars defined in Fig.~\ref{2C} plotted in green and magenta
        colors, respectively.  In the lower panels we have superposed on
        these CMDs the MSa fiducial line (green continuous line) and the
        MSb fiducial (magenta dashed line),
calculated in each case by the median-color, spline, sigma-clip
procedure that was described earlier in the text. }
         \label{MScriteria}
   \end{figure*}

With high-accuracy photometric measurements in nine bands, 36 different
CMDs could be generated, if we were to use all possible combinations of
magnitude and color.
In the upper panels of Figure \ref{MScriteria} we show three of the most
representative of these CMDs, zoomed around the MS region, with each
star colored green or magenta, according to its membership in MSa or MSb
as shown in Fig.~\ref{2C}.
In the bottom panels of Fig.~\ref{MScriteria} the same CMDs are
replicated, and superimposed on the observed CMDs are the fiducial ridge
lines of MSa and MSb, derived for each CMD using the method described
above.

\begin{figure}[ht!]
   \centering
   \epsscale{0.75}
   \plotone{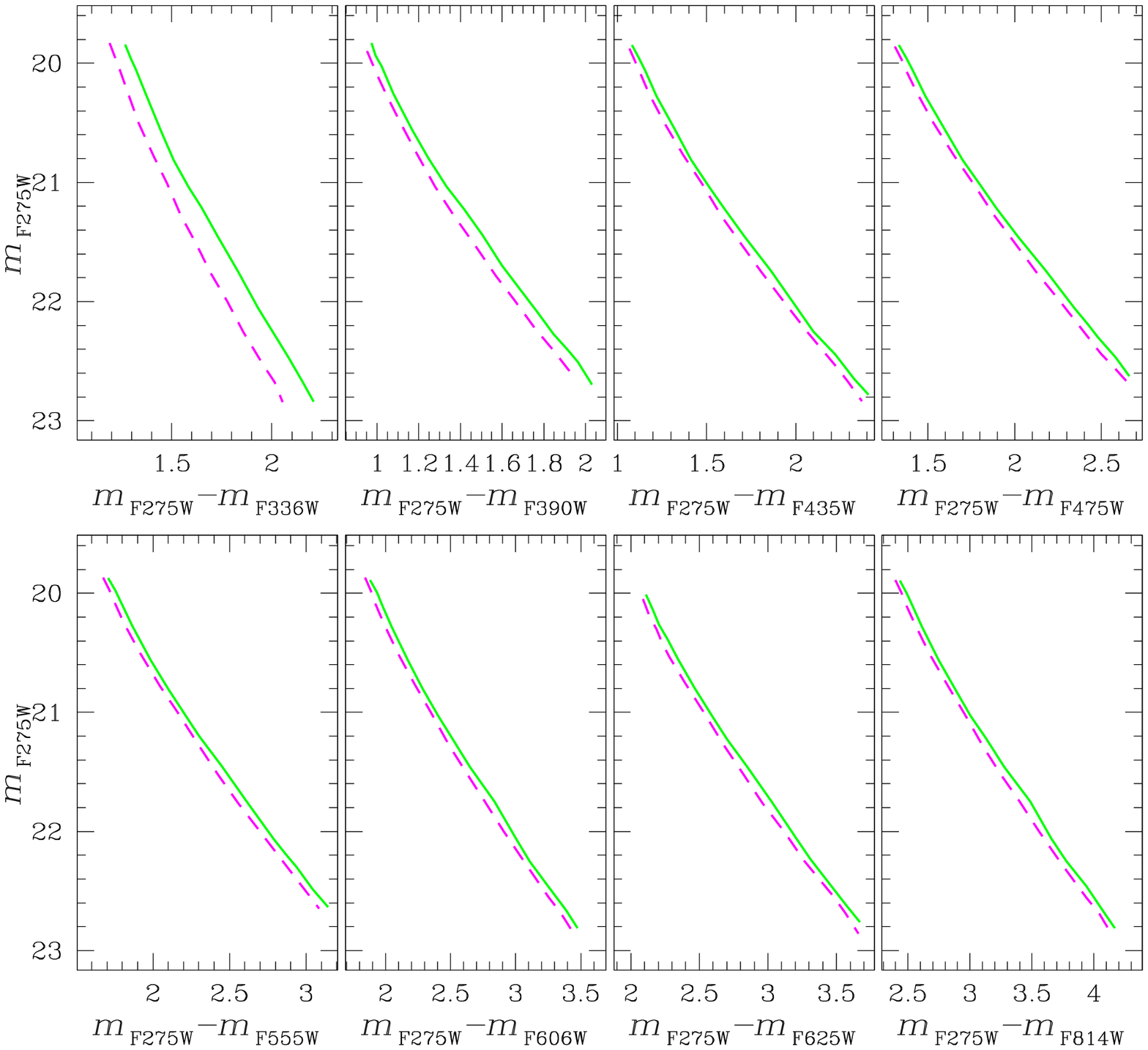}
      \caption{ MS fiducial lines for MSa (green continuous line)
        and MSb (magenta dashed line) in the $m_{\rm F275W}$ vs.\
        $m_{\rm F275W}-m_{\rm X}$ plane. In all these combinations of
        magnitude and colors the MSa stars are systematically redder
        than the MSb stars.}
         \label{fid275}
   \end{figure}

\begin{figure}[ht!]
   \centering
   \epsscale{0.75}
   \plotone{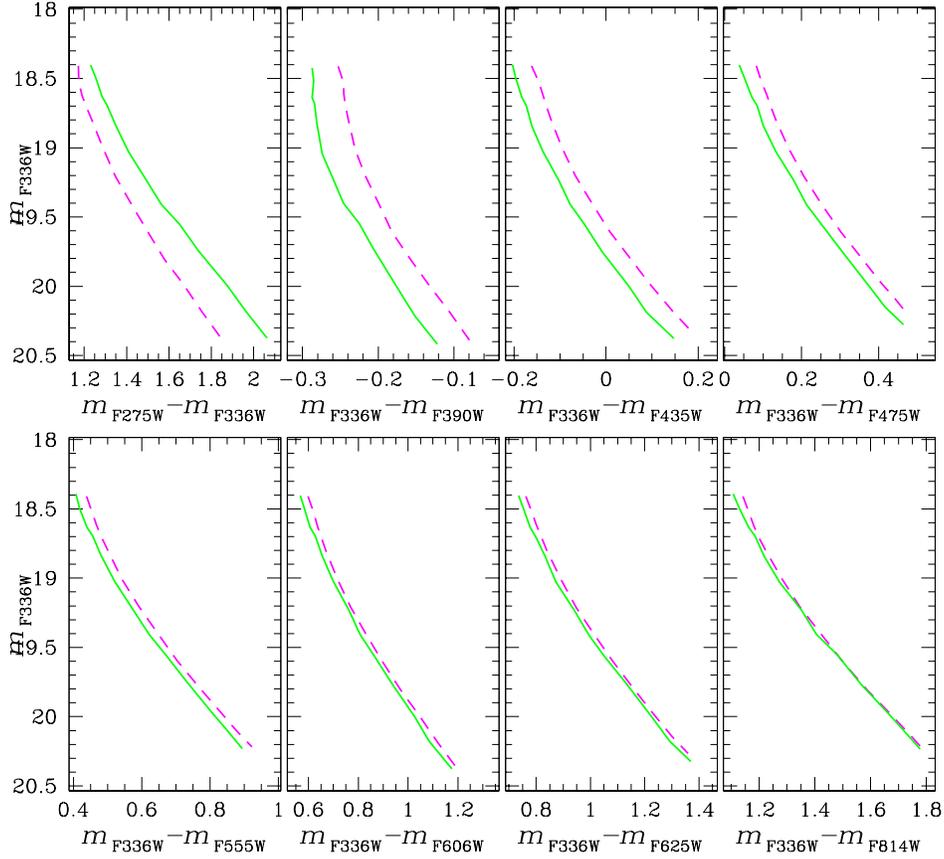}
      \caption{ MS fiducial lines for MSa (green continuous line) and
        MSb (magenta dashed line) in the $m_{\rm F336W}$ vs.\ $m_{\rm
          F336W}$-$m_{\rm X}$ ($m_{\rm F275W}-m_{\rm F336W}$ in the
        upper left panel) plane. As in all the CMDs of
        Fig.~\ref{fid275}, MSb is bluer than MSa in all the $m_{\rm
          F336W}$ vs.\ $m_{\rm F275W}$-$m_{\rm F336W}$ CMDs.  But
        notably, MSa becomes bluer than MSb in the
$m_{\rm F275W}-m_{\rm F336W}$ color.} 
         \label{fid336}
   \end{figure}

\begin{figure}[ht!]
   \centering
   \epsscale{0.75}
   \plotone{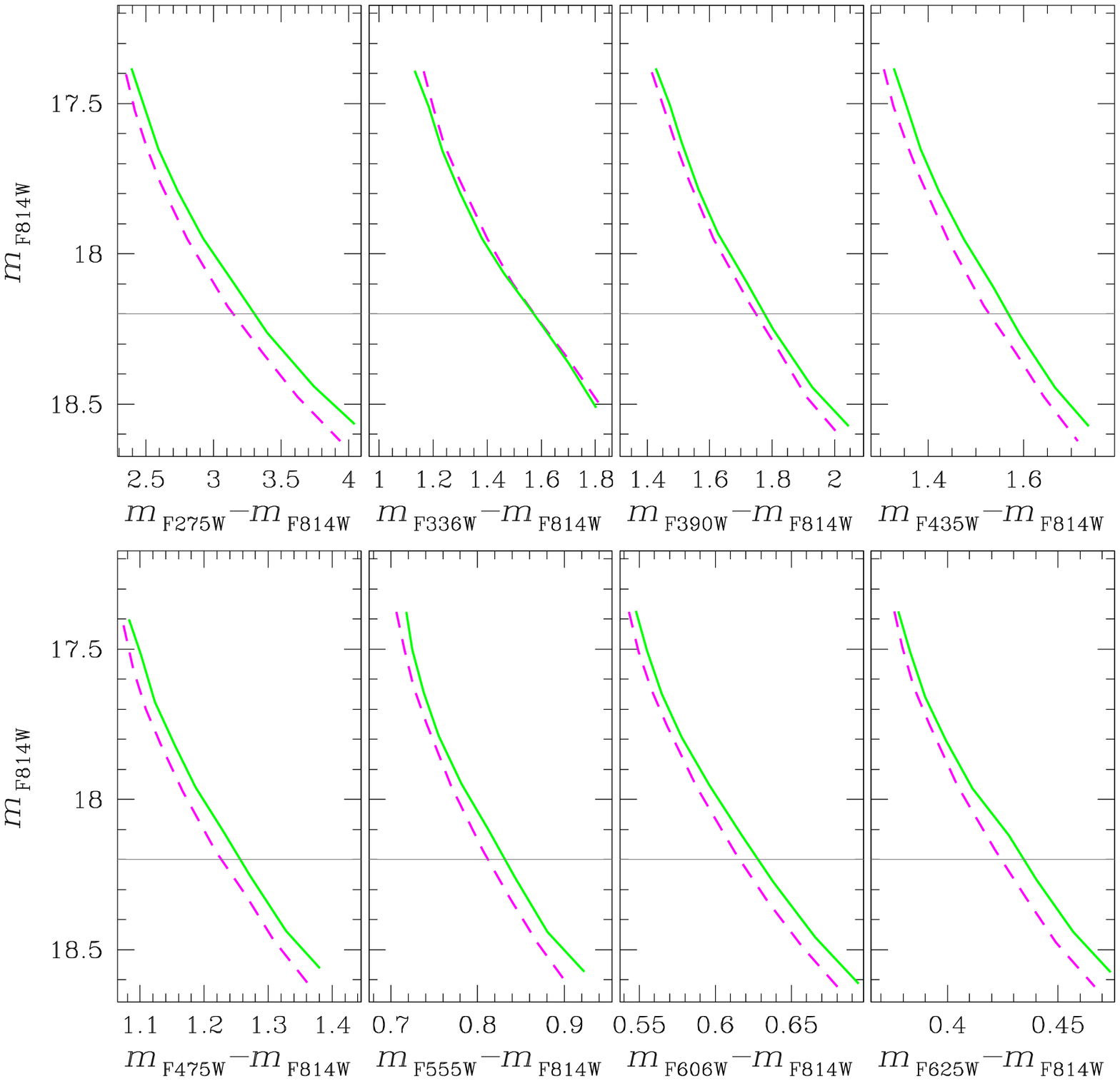}
      \caption{ MS fiducial lines for MSa (green continuous line) and
        MSb (magenta dashed line) in the $m_{\rm F814W}$ vs.\ $m_{\rm
        X}$-$m_{\rm F814W}$ plane. In these combinations of magnitude
        and color the MSa stars are systematically redder than the MSb
        stars, apart from the $m_{\rm F814W}$ vs.\ $m_{\rm F336W}-m_{\rm
        F814W}$ CMD, where MSb is marginally redder than MSa. 
          Horizontal grey lines mark the magnitude at which we have
      calculated the color difference between the two MSs ($m_{\rm
        F814W}$=18.2, see text for more details). }
         \label{fid814}
   \end{figure}

The fiducials for the two MSs are then shown again in Fig.~\ref{fid275}
($m_{\rm F275W}$ vs.\ $m_{\rm F275W}-m_{\rm X}$), Fig.~\ref{fid336}
($m_{\rm F336W}$ vs.\ $m_{\rm F336W}-m_{\rm X}$ or $m_{\rm X}-m_{\rm
F336W}$), and Fig.~\ref{fid814} ($m_{\rm F814W}$ vs.\ $m_{\rm
  X}-m_{\rm F814W}$), where $m_{\rm X}$ is indicated on the ordinate of each of row
of figures. We note that MSa is redder than MSb in most of these CMDs,
but the color ordering of the two fiducial lines 
is otherwise in some of them:
MSa stars become bluer than MSb stars in the $m_{\rm F336W}-m_{\rm X}$
colors (Fig.~\ref{MScriteria}, lower right panel, and
Fig.~\ref{fid336}).


\section{Interpreting the Two Branches of the Main Sequence}
\label{YcnoMS}

Surely the bizarre behavior of the two branches of the MS in our various
CMDs is telling us something about the physical origin of the
split.  
  In hopes of clarifying the complicated observational picture, we
  will focus on a critical subset of the data, then we will 
    compare the observations against synthetic spectra that have been
    calculated for atmospheres with various chemical compositions.

\begin{figure*}[ht!]
   \centering
   \epsscale{0.75}
   \plotone{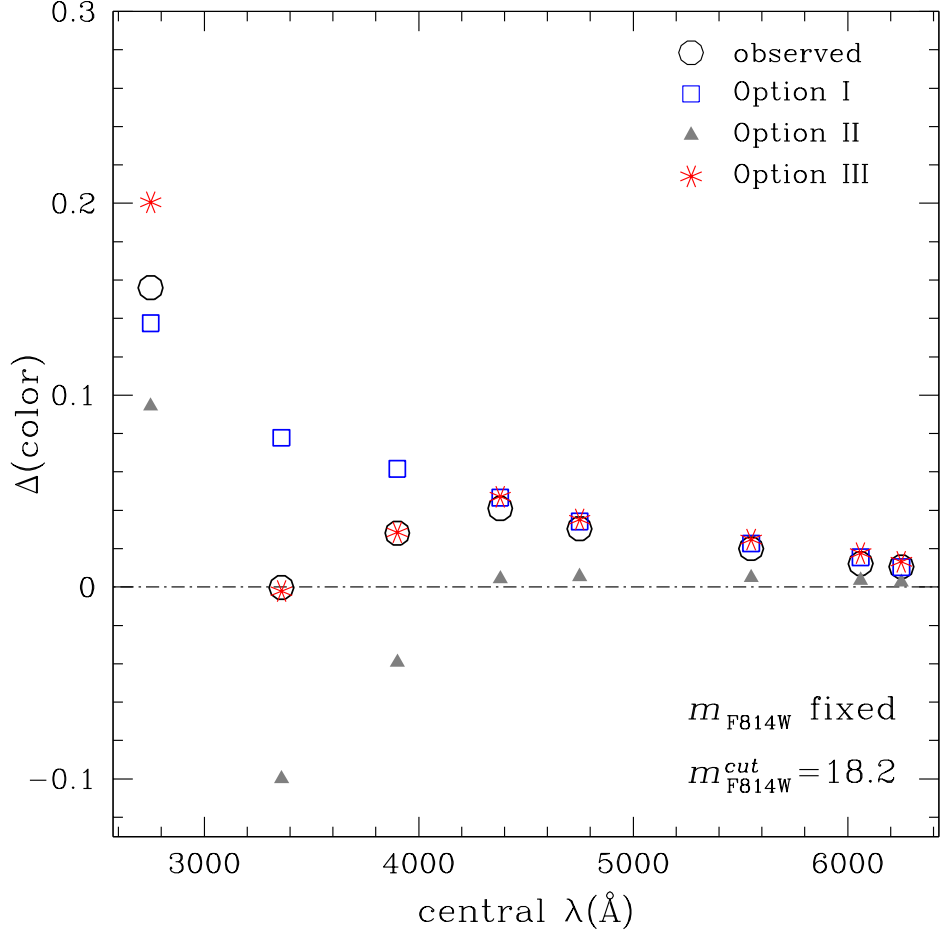}
      \caption{Color separations of the ridge lines of MSa and MSb for
        different color baselines.  The colors used are $m_{\rm
        X}-m_{\rm F814W}$, where X is one of the other eight filters.
        Each color separation is shown at the central wavelength of its
        filter X.  (For example, the leftmost point is the separation of
        the two sequences in $m_{\rm F275W}-m_{\rm F814W}$.)  All color
        separations are measured at $m_{\rm F814W}=18.2$.  A positive
        value of $\Delta$(color) means that MSa is bluer; negative would
        mean that MSa is redder.  Observations are plotted as open
        circles, while the color differences expected from theoretical
        Options I, II, and III are shown as blue squares, gray
        triangles, and red asterisks, respectively.}
         \label{AT12}
   \end{figure*}

The challenge of choosing a critical subset from our mass of
observational results is similar to the one faced by Bellini et
al.\ (2010) in their multicolor study of $\omega$ Centauri; we follow
their lead, and look at the separation of the sequences as a function of
the color index in which they are observed.  Figure \ref{AT12} shows the
separation of MSa and MSb in each of eight colors; the figure also
compares these observational separations of the sequences with
calculated separations, from three theoretical scenarios that we will
describe below.

On the theoretical side, multiplicity of MSs has been attributed to
differences in helium content (e.g., Norris 2004, D'Antona et al.\ 2005,
Piotto et al.\ 2005, 2007) or in light-element abundances (e.g.,
Sbordone et al.\ 2011), so we will explore both of these possibilities,
by making three different choices for the abundances of He, C, N, and O.
We will calculate synthetic colors for MSa and MSb stars for each of
these options, seeking the composition that best reproduces all the
observed color differences between the two MSs.
All three of the options use the same abundance mixture for MSa. For He
in MSa we choose the primordial He abundance, $Y=0.256$, and, following
Cannon et al.\ (1998), we choose [C/Fe] = 0.06 and [N/Fe] = 0.20, which
are typical for CN-weak stars.  As for [O/Fe], we take 0.40, typical for
first-generation stars (Carretta et al.\ 2009a,b).  For MSb stars we try
three different options.  In Option I we assume that helium is the only
cause of the MS split, and adopt $Y=0.28$.  In Option II we keep helium
the same in the two populations but instead change the light-element
abundances, adopting the values listed in Table 2.  Finally, in Option
III we adjust both helium and the light elements, again as given in
Table 2.

We chose to characterize the fiducial sequences by measuring the
color difference between MSa and MSb at the reference magnitude of
$m_{\rm F814W} = 18.2$, that is, the magnitude level indicated by
  the horizontal lines in Fig.~\ref{fid814}.  For an assumed distance modulus 
$(m-M)_{\rm F814W}=13.41$  and a reddening $E(B-V)=0.04$, 
this corresponds to an absolute magnitude of $M_{\rm F814W}=4.85$. 
 The adopted distance modulus and reddening are those that provide
the best fitting of the isochrones to 
the data in the 
$m_{\rm F606W}$ vs.\ $m_{\rm F606W}-m_{\rm F814W}$ plane, and are in
agreement with those provided by Gratton et al.\ (2003) and Harris
(1996, Dec.\ 2010 update ). To determine the absorption in the  F814W band we used the
relations given by Bedin et al.\ (2005). We assumed 
[Fe/H] = $-$0.75 and [$\alpha$/Fe] = 0.4,
in agreement with the values found by Carretta et
al.\ (2009).
To compare the observations against synthetic photometry, we adopted
the isochrones from the Teramo group (Pietrinferni et al.\ 2004, 2006)
 specifically calculated
for the populations listed in Table 2, then determined $T_{\rm eff}$
and $\log g$ for MS stars at $M_{\rm F814W}=4.85$.
These temperatures and gravities were then used to
calculate model atmospheres with the ATLAS12 code (Kurucz 2005, Castelli
2005, Sbordone 2005, Sbordone et al.\ 2007), which allows us to use
arbitrary chemical compositions. Spectral synthesis from 2000 \AA\ to
10000 \AA\ was then performed using the SYNTHE code (Kurucz 2005,
Sbordone et al.\ 2007), and
the resulting synthetic spectra were integrated over the transmission
of each of our nine filters to produce the synthetic magnitudes and colors.

We did this separately for an MSa star and for an MSb star, using for
the latter each
of the three composition options that Table 2 lists for MSb.  We
concentrate here on the eight $m_{\rm X}-m_{\rm F814W}$ colors,
comparing the synthetic MSa $-$ MSb color differences with the observed
ones shown in Figure~\ref{AT12}.
The results of our synthetic-spectrum calculations for Option I are
shown in the figure as blue squares.  At wavelengths longer than
$\sim 4300$ \AA\ there is good agreement with the observed color
differences between the two MSs, but sizable discrepancies appear at
shorter wavelengths, in particular in the F336W band, and we conclude
that helium alone cannot account for the observed MS split.  The colors
that result from Option II (same helium but different CNO proportions)
are plotted as gray triangles in Figure~\ref{AT12}.  Again there is a
strong discrepancy between the simulated and the observed color
differences in the two near-UV bands (F336W and F390W), but now in the
opposite direction.  Turning instead to Option III, with differences in
both helium and the CNO elements, we see that the red asterisks in
Figure~\ref{AT12} are in fair agreement with all of the observed color
differences.  The agreement is somewhat poorer for the F275W filter, but
note that we have not fine-tuned the composition differences between the
two MSs, and that UV colors can be very sensitive to the adopted mix of
He and the CNO elements.

\begin{table}[ht!]
\begin{center}  
\scriptsize {
\caption{Parameters used to simulate synthetic spectra of an MSa and
an MSb star with $m_{\rm F814W}$=18.2, for the three assumed options.
  For all the populations we assumed [Fe/H] = $-$0.75 and $[\alpha/Fe]$ = 0.4. }

\begin{tabular}{ccccccc}
\hline
\hline
MS (Option) & $T_{\rm eff}$ & log g & $Y$  & [C/Fe] & [N/Fe] & [O/Fe] \\
\hline
MSa (all)   &  5563         & 5.42 & 0.256 &   0.06 & 0.20   &   0.40 \\
MSb (I)     &  5648         & 5.41 & 0.288 &   0.06 & 0.20   &   0.40 \\
MSb (II)    &  5563         & 5.42 & 0.256 &$-$0.15 & 1.05   &$-$0.10 \\        
MSb (III)   &  5592         & 5.41 & 0.272 &$-$0.15 & 1.05   &$-$0.10 \\         
\hline
\hline
\label{paramMSs}
\end{tabular}
}
\end{center}
\end{table}

%
   \begin{figure}[ht!]
   \centering
   \epsscale{0.75}
   \plotone{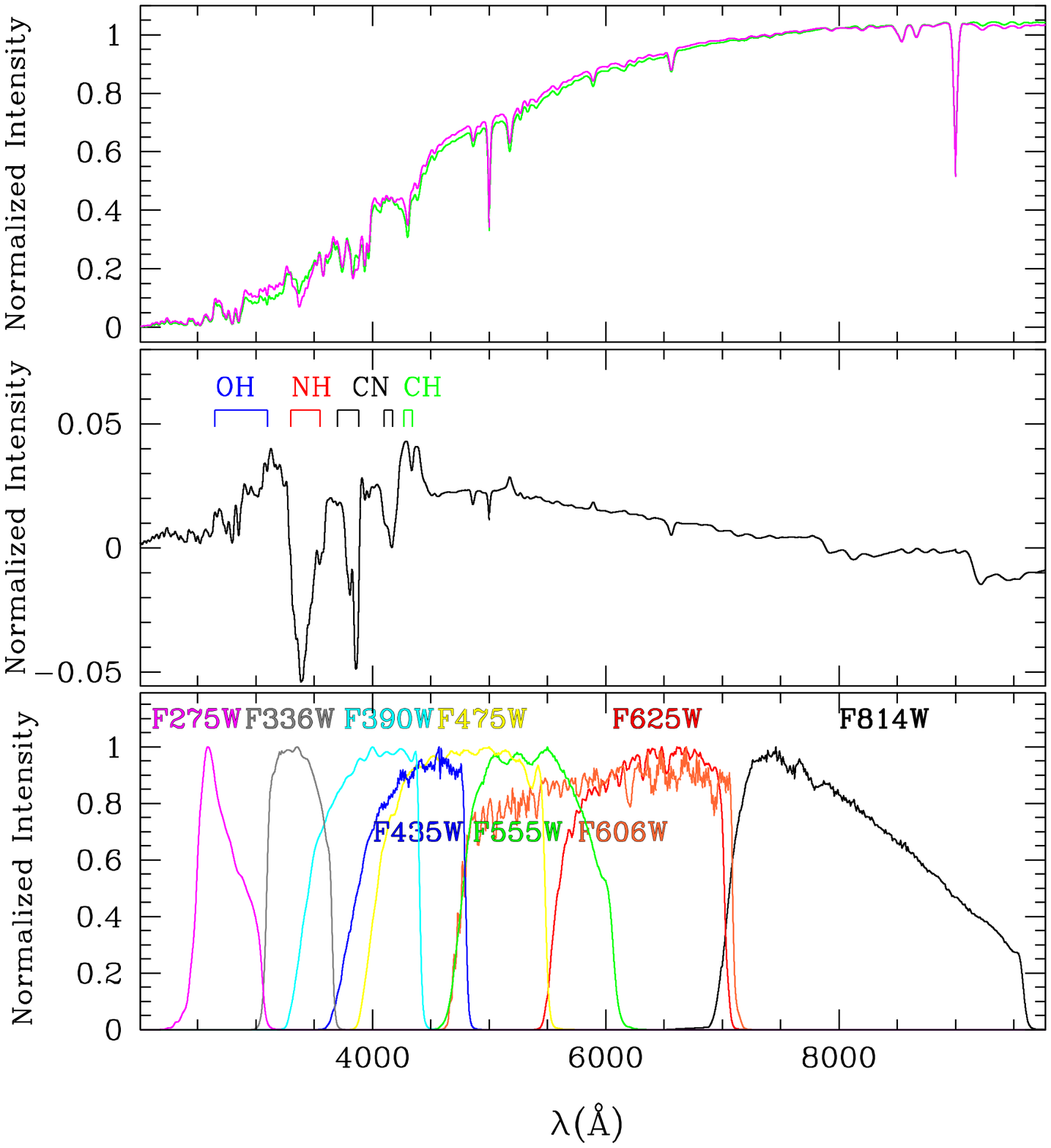}
      \caption{\textit{Upper panel:} comparison of the synthetic spectra
        of an MSa star (green) and an MSb star (magenta). See text for
        more details. \textit{Middle panel:} Difference between the
        spectra of the MSa star and the MSb star. \textit{Lower panel:}
        Normalized responses of the {\it HST} filters used in this
        paper. }
         \label{spettri12}
   \end{figure}
%

The effect of chemical composition differences on MS colors can be
summarized as follows.  The main effect of increased helium is to
increase the temperatures, making all colors bluer. The effects of CNO
are more subtle, since a specific molecule affects some bands but not
others. Thus nitrogen affects the near-UV F336W and F390W bands via the
NH band and CN bands, respectively, whereas oxygen affects the F275W
band, via the OH band.  This is illustrated in the top panel of
Figure~\ref{spettri12}, showing the synthetic spectra of an MSa and an
MSb star as calculated for Option III (in green and magenta
respectively).  The difference between the two is shown in the middle
panel, while the band-passes of our filters are plotted in the bottom
panel.

These results indicate that the observed color differences between MSa
and MSb can be understood by assuming that MSa corresponds to a first
stellar generation, with primordial He, and O-rich/N-poor stars, whereas
MSb corresponds to a population that is enriched in He and N but
depleted in O.  This need for differences in both helium and CNO to
account for all the color differences fits quite well with
nucleosynthesis expectations, as helium-enriched stellar regions are
also inevitably oxygen-depleted and nitrogen-enriched, and vice versa
--- for example, in the layers subject to the {\it second dredge-up}, at
the beginning of the AGB phase, or those subject to the so-called
envelope-burning process, later during this phase (Renzini \& Voli 1981,
Ventura \& D'Antona 2009).

%
\section{Multiple stellar populations along the Subgiant Branch}
\label{sec:SGB}

The first photometric evidence for multiple populations in 47 Tuc came
from the discovery of its bimodal SGB (Anderson et al.\ 2009).  These
authors analyzed the large number
of ACS/WFC archival {\it HST} images of the core of 47 Tuc, and found
that the SGB is split into two distinct components: a
brighter SGB showing an intrinsic broadening in the F475W magnitude, and
a second one about 0.05 mag fainter, containing a minor fraction (a few
per cent) of the stars.  Anderson et al.\ also examined the SGB in an
outer field, $\sim 6$ arcmin west of the cluster center, finding a
similar vertical spread of the SGB.  These SGBs have been further
investigated by Piotto et al.\ (2011), with a multi-band photometric
analysis, showing that the magnitude difference between the faint and
bright SGBs is almost constant in each of the nine {\it HST} filters
they used, and finding a similar behavior in five other Galactic GCs
that have multiple SGBs.  Piotto et al.\ also found that the faint SGB
makes up $8\pm3$ per cent of the stars in the central field of 47 Tuc.
Di 
Criscienzo et al.\ (2010) suggest that the spread in luminosity of the
bright SGB can be accounted for by a small spread in helium ($\Delta Y
\sim 0.02$), whereas they suggest that the minor population that makes
up the faint SGB (fSGB) would be characterized by a small increase in
C+N+O.

   \begin{figure*}[ht!]
   \centering
   \epsscale{0.75}
   \plotone{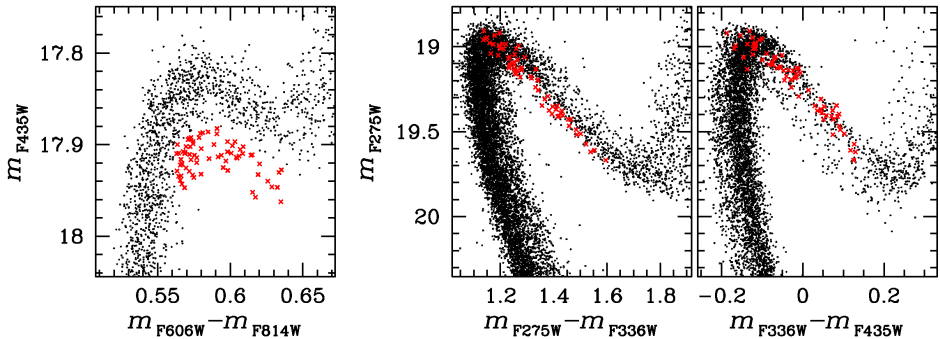}
      \caption{ CMDs zoomed around the SGB: $m_{\rm F435W}$ vs.\ $m_{\rm F606W}$-$m_{\rm F814W}$
        (left panel), $m_{\rm F275W}$ vs.\ $m_{\rm F275W}-m_{\rm F336W}$
        (middle panel) and $m_{\rm F275W}$ vs.\ $m_{\rm F336W}-m_{\rm  F435W}$
          (right panel). The leftmost
        CMD has been used to select a sample of faint SGB stars,
        represented by red X's in all three CMDs.}
         \label{3SGBs}
   \end{figure*}
%

In our multicolor set of CMDs, the SGB region turns out to be
unexpectedly complex.  Figure~\ref{3SGBs} shows in the left panel the
visible-light CMD of $m_{\rm F435W}$ vs.\ $m_{\rm F606W}-m_{\rm F814W}$,
in which the fSGB component of Anderson et al., marked in red, is
clearly visible as a separate sequence. In the right panels are the
ultraviolet CMDs, with $m_{\rm F275W}$ vs.\ $m_{\rm F275W}-m_{\rm F336W}$
and $m_{\rm F275W}$ vs. $m_{\rm F336W}-m_{\rm F435W}$; in them the same
stars instead fall on top of the numerically dominant sequence, which in
turn splits into two separate sequences, bringing the total to three
recognizable SGBs.  How can this be?

By analogy with what we did for the MS stars, in Figure~\ref{2Csgb} we
plot the UV two-color diagram of the SGB stars.  Here again we see a
bimodal distribution, and we therefore repeat the points on the right,
and draw a separator line.  Again by analogy, we color the stars on
either side of the line green and magenta, respectively, and attach to
the two regions the names SGBa and SGBb.  In this case, however, we also
distinguish with red X's the stars that belonged to the fSGB in the
left panel of Fig.~\ref{3SGBs}.

The next step is to bring to bear on the SGB puzzle our full multicolor
resources. 
In Figure \ref{SGBs} we show a 4 $\times$ 4 array of CMDs, with a
different magnitude in each row, plotted against a different color in
each column.  Although these CMDs are worth looking at one by one, taken
together they add up to an {\it embarras de richesse}.  We therefore
note for the reader the characteristics that strike us as systematic and
significant:

   \begin{figure}[ht!]
   \centering
   \epsscale{0.75}
   \plotone{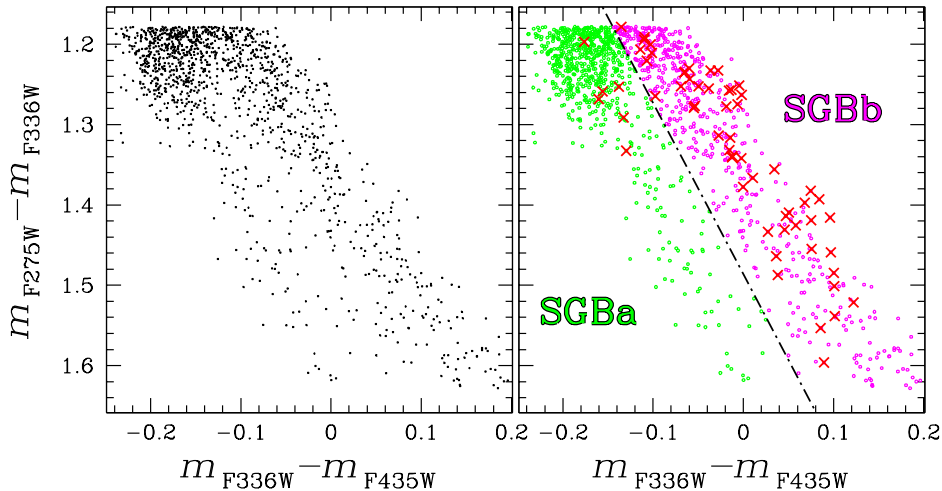}
     \caption{$ m_{\rm F275W}-m_{\rm F336W}$ vs.\ $m_{\rm
         F336W}-m_{\rm F435W}$ two-color diagram for SGB
       stars. The dot-dash line in the right panel arbitrarily
       separates the SGBa and SGBb stars, which are plotted green and
       magenta, respectively.}
        \label{2Csgb}
  \end{figure}

   \begin{figure*}[ht!]
   \centering 
   \epsscale{0.75}
   \plotone{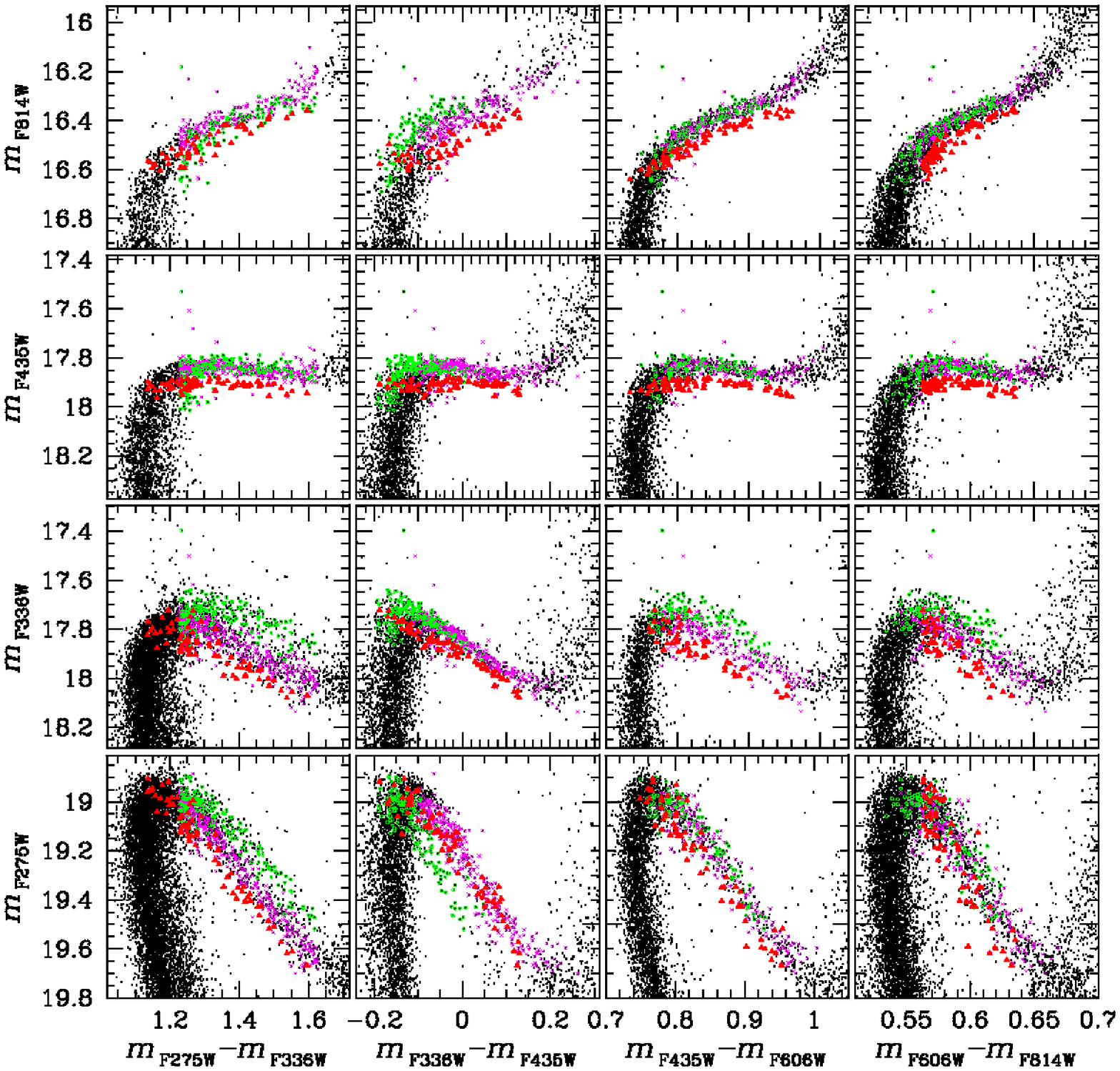}
      \caption{ Collection of CMDs from {\it HST} photometry, zoomed
        around the SGB. The two groups of SGBa and SGBb stars defined
        in Fig.~\ref{2Csgb} are plotted in green and magenta, respectively, while
        we have used red symbols for the faint SGB stars selected
      in Fig.~\ref{3SGBs}.}
         \label{SGBs}
   \end{figure*}
%

(1) SGBa stars share some similarities with MSa:\ in the CMDs that use
   the $m_{\rm F275W}$-$m_{\rm F336W}$ color they are on average redder
   than the other SGB stars, but when the $m_{\rm F336W}-m_{\rm F435W}$
   color is used instead, the SGBa stars are on average bluer than the
   bulk of SGB stars.  No significant color difference between the two
   SGB groups is evident in the other color combinations.

(2) SGBa stars are typically brighter than SGBb stars in $m_{\rm F336W}$
   (note in particular the third row of panels). Their being brighter
   than others in this band explains why they are redder than SGBb and
   fSGB in the $m_{\rm F275W}-m_{\rm F336W}$ color, and bluer than
   them in the $m_{\rm F336W}-m_{\rm F435W}$ color. No major
   systematic magnitude difference between SGBa and the two other SGB
   groups is apparent in any of the other panels.

(3) In all of the CMDs the fSGB stars follow the same trend as the bulk
   of the SGBb stars, with
   the fSGB running parallel to SGBb.  
 
The nature of two of the three SGBs is now clear.  In the first and
second CMDs in the bottom row of Fig.~\ref{SGBs} ($m_{\rm F275W}$
vs.\ $m_{\rm F275W}-m_{\rm F336W}$ and $m_{\rm F275W}$ vs.\ $m_{\rm
  F336W}-m_{\rm F435W}$) the green and magenta sequences interchange
their order, exactly as we saw for MSa and MSb in these same colors
(upper left and upper right CMDs in Fig.~\ref{MScriteria}).  This justifies
the name that we gave this pair of sequences, SGBa and
SGBb, since they appear to be the direct descendants of MSa and MSb.

The sequence that we have marked in red establishes the existence of a
third population in 47 Tuc, but we see it only here on the SGB; in other
parts of the CMD we find no evidence of this third sequence.  It is ironic
that the very stars whose existence led us to study multiple populations
in 47 Tuc should be left as an anomaly that does not fit into the
picture that will emerge at the end of the present paper. 

%
   \begin{figure*}[ht!]
   \centering
   \epsscale{0.95}
   \plotone{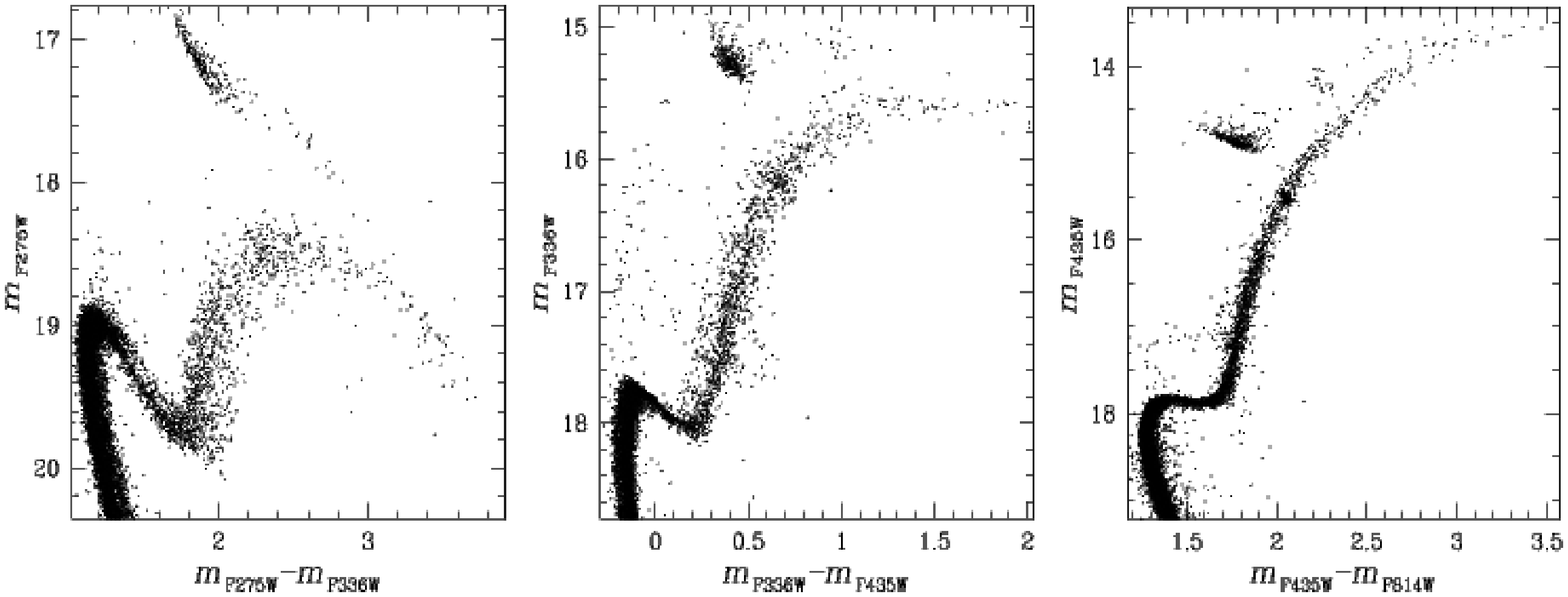}
      \caption{This figure illustrates the three CMDs from {\it HST}
        photometry that best summarize the behavior of RGB stars. Note
        the wide color spread of $\sim$ 0.1--0.2 mag in the left and the
        middle CMDs, which also show some hints of color bimodality. 
       The RGB becomes narrower and well defined in the right-hand CMD.}
         \label{RGBshst}
   \end{figure*}
%

\section{Multiple stellar populations on the red giant branch}
\label{sec:RGB}

Figure \ref{RGBshst} shows the red-giant region of 47 Tuc in three
different CMDs.  Both the $m_{\rm F275W}$ vs.\ $m_{\rm F275W}$-$m_{\rm
F336W}$ and the $m_{\rm F336W}$ vs.\ $m_{\rm F336W}-m_{\rm F435W}$
diagrams suggest that the RGB is intrinsically broad; the observed color
spread of 0.1--0.2 mag is much larger than the photometric error, which
for these bright RGB stars is less than 0.02 mag in color, and there is
even some hint of a double color distribution.  However, in the $m_{\rm
F435W}$ vs.\ $m_{\rm F435W}-m_{\rm F814W}$ CMD (rightmost panel) the RGB
is quite narrow.

%
   \begin{figure*}[ht!]
   \centering
   \epsscale{0.75}
   \plotone{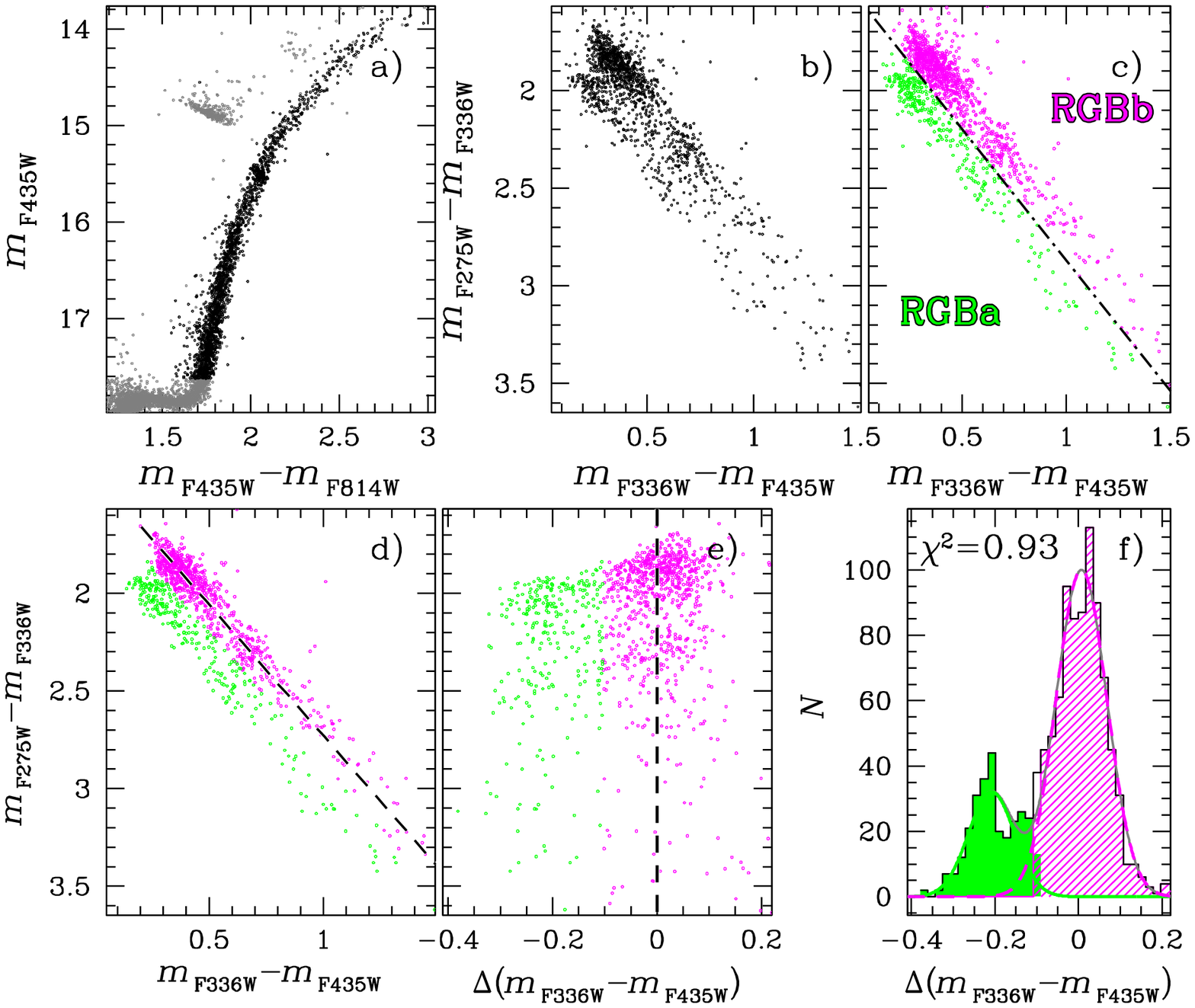}
%
      \caption{\textit{Panel a}: $m_{\rm F435W}$ vs.\ $m_{\rm
          F435W}-m_{\rm F814W}$ CMD from ACS/WFC photometry. The black
        points are the RGB stars that we show in the other panels.
        \textit{b}: $m_{\rm F275W}-m_{\rm F336W}$ vs.\ $m_{\rm
          F336W}-m_{\rm F435W}$ two-color diagram for these stars.
        \textit{c}: We arbitrarily drew by hand the dash-dot line that
        separates RGBa (green) and RGBb (magenta).  \textit{d}: The
        dashed line is the fiducial sequence of the RGBb stars, drawn by
        the fitting method that we described in Sect.\ \ref{sec:MS}.
        \textit{e}: The verticalized colors of the stars.  \textit{f}:
        Histogram of the colors in panel \textit{e}, with a fit by two
        Gaussians, whose sum is shown by the solid gray line.} 
         \label{RGBhst2C}
   \end{figure*}
%
To investigate this behavior further, we again turn to the $m_{\rm
  F275W}-m_{\rm F336W}$ vs.\ $m_{\rm F336W}-m_{\rm F435W}$ two-color
diagram.  In panel {\it a} of Figure~\ref{RGBhst2C} we show a zoomed
CMD, in which the RGB stars are indicated in black.  Panel {\it b}
shows the two-color diagram of those stars.  The distribution of stars
in the diagram 
is clearly bimodal, and in panel {\it c} we arbitrarily draw a line
separating two groups of stars, which we name RGBa and RGBb, again
color-coded green and magenta, respectively.
The lower panels of the figure show the steps by which numerical
fractions are derived for the RGBa and RGBb populations.  In accordance
with the procedure introduced in Sect.\ 3, we derived a fiducial line
for RGBb in the two-color diagram; the line is shown in panel {\it d}.
In panel {\it
 e} we have subtracted from the color of each star the color of the
fiducial line at the magnitude of the star, so as to verticalize the
sequences.  In panel {\it f} we plot a histogram of the verticalized
colors, and fit it with Gaussians to represent the RGBa and RGBb
populations.
The result is that RGBa represents $19\pm3$\% of the total.

\subsection{The double RGB in the outer parts of the cluster}

\begin{figure}[ht!]
   \centering
   \epsscale{0.75}
   \plotone{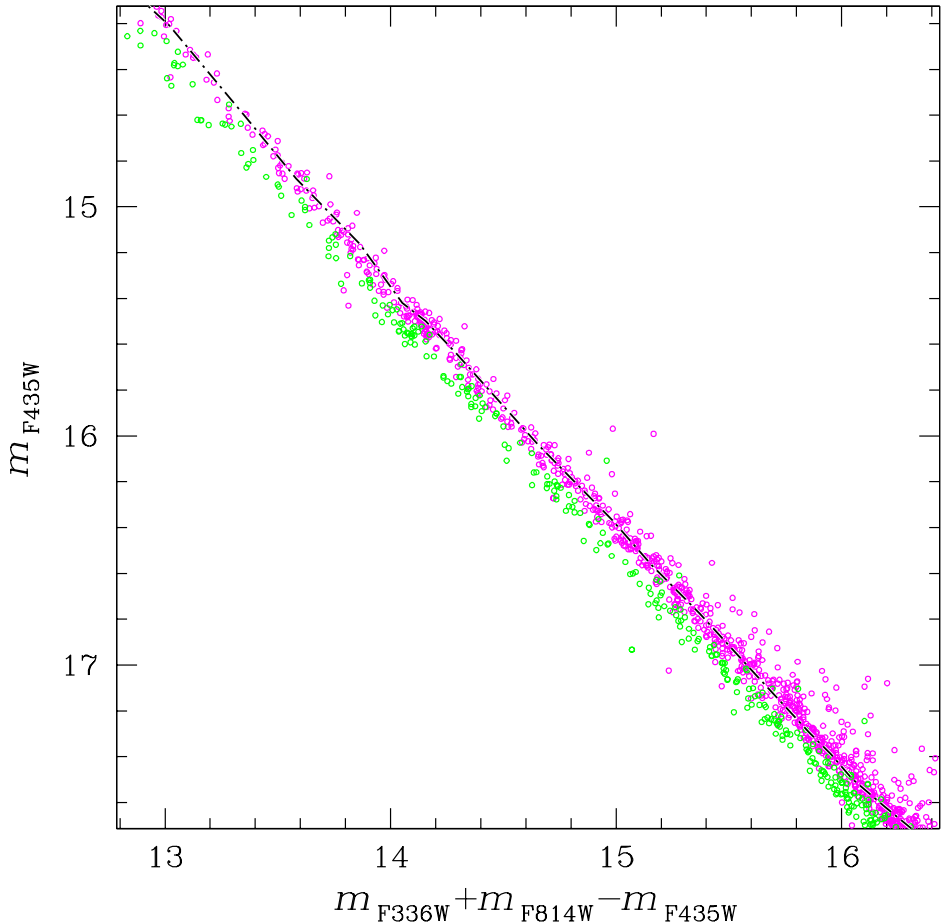}
      \caption{The $m_{\rm F435W}$ vs.\ $m_{\rm F336W}+m_{\rm
          F814W}-m_{\rm F435W}$ diagram for RGB stars. The dash-dot
          black line is the fiducial line of the RGBb sequence.}
         \label{RGBhst3Ma}
   \end{figure}

Ground-based photometry available for a field of view that
goes out to 25 arcmin allows us to investigate the radial behavior of
the populations.  However, the lack of an equivalent to the F275W
passband deprives us of our sharpest tool, and we must look for the best
population discriminant among the ground-based passbands that are
available.  Fortunately, ground-based $U$, $B$, $V$, and $I$ are quite
similar to our {\it HST} F336W, F435W, F555W and F814W bands.  Our
procedure, then, was as follows: Since our use of the F275W and F336W
bands had separated the RGBa and RGBb stars, we took those two samples
of stars and examined their behavior in the many different diagrams that
can be plotted using only the four bands that have ground-based
equivalents, in order to see if we could find a surrogate for the two
missing bands.  As a result, we found that
RGB stars separate into two groups when we plot $m_{F435W}$ against the
combination $m_{\rm F336W}$+$m_{\rm F814W}-m_{\rm F435W}$, as is shown
in Figure \ref{RGBhst3Ma}.

%
%

\begin{figure}[ht!]
   \centering
   \epsscale{0.75}
   \plotone{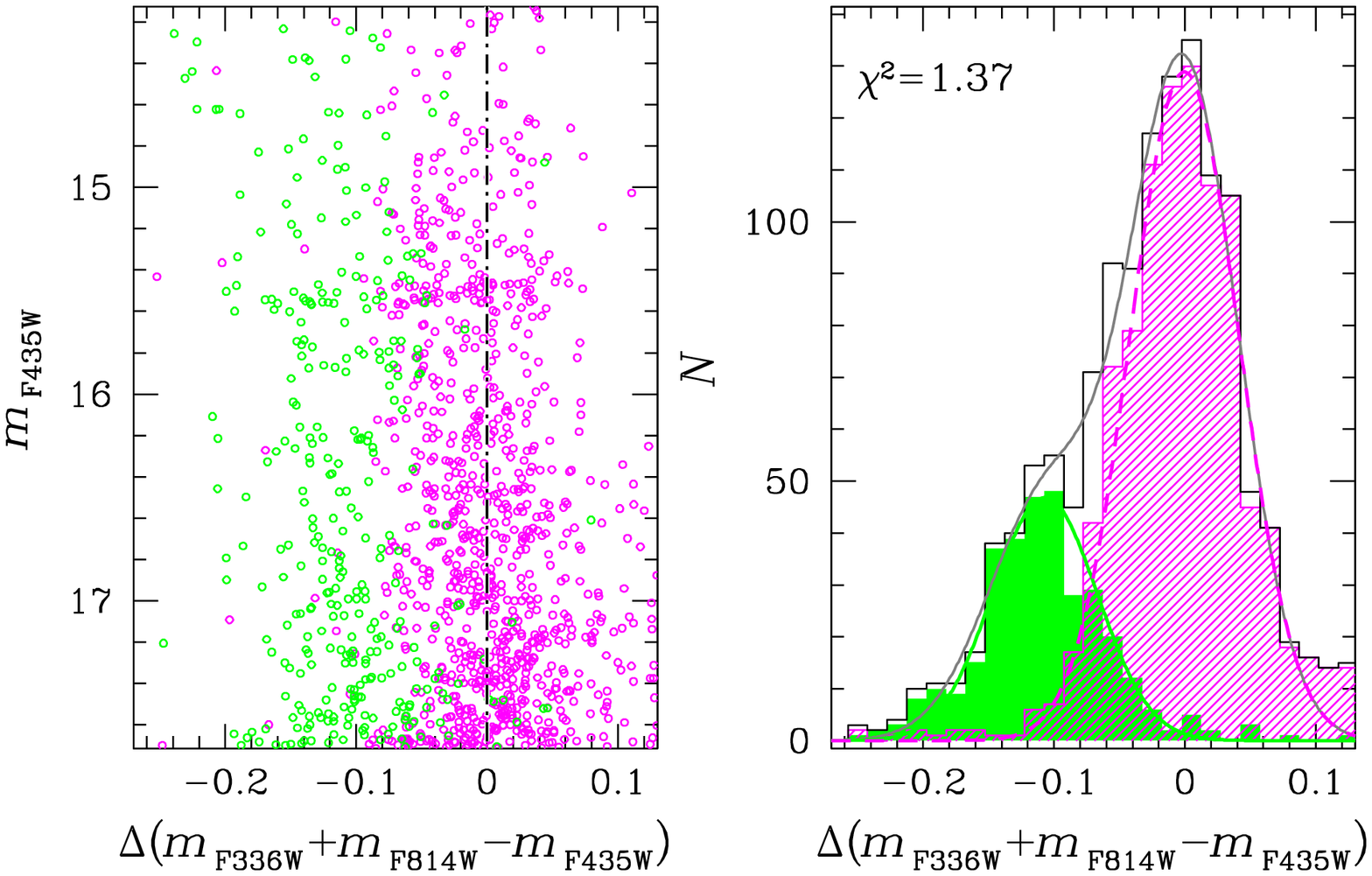}
      \caption{\textit{Left}: The same points as in
          Fig.~\ref{RGBhst3Ma}, but after subtracting from the $m_{\rm
          F336W}+m_{\rm F814W}-m_{\rm F435W}$ value of each star the
          value of the fiducial at the same magnitude.
        \textit{Right}: Histogram of the colors in the left panel. The
        gray line is the best fit of two Gaussians, which are plotted as
        dotted and dashed black lines, and are shaded green and magenta
        respectively.}
         \label{RGBhst3Mb}
   \end{figure}

Just to demonstrate the effectiveness of this separation, we use it to
separate the stars of RGBa and RGBb, instead of
using the easier route offered by Figure~\ref{RGBhst2C}.  As in similar
situations encountered earlier,
we draw a fiducial line for RGBb, by putting a spline through the median
colors in successive short intervals of magnitude, and doing an iterated
sigma-clipping of outliers.  We again subtract the color of the fiducial
line from the color of each star, to produce the verticalized colors
shown in the left panel of Figure \ref{RGBhst3Mb}.  The
corresponding two-Gaussian fit to the color distributions of the
resulting RGBa and RGBb
stars
is shown in the right panel of the figure.

%

   \begin{figure*}[ht!]
   \centering
   \epsscale{0.4}
   \plotone{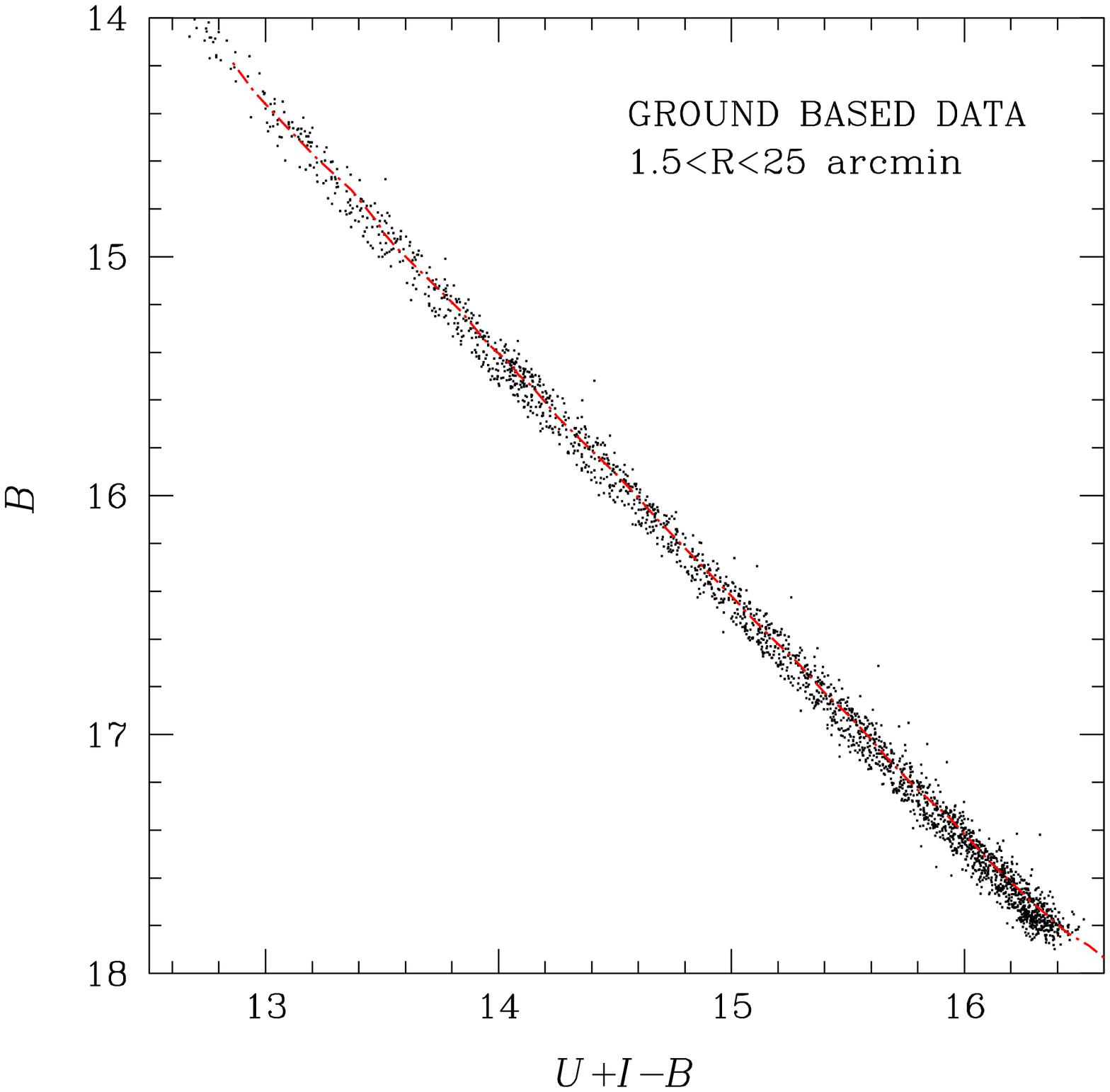}
   \plotone{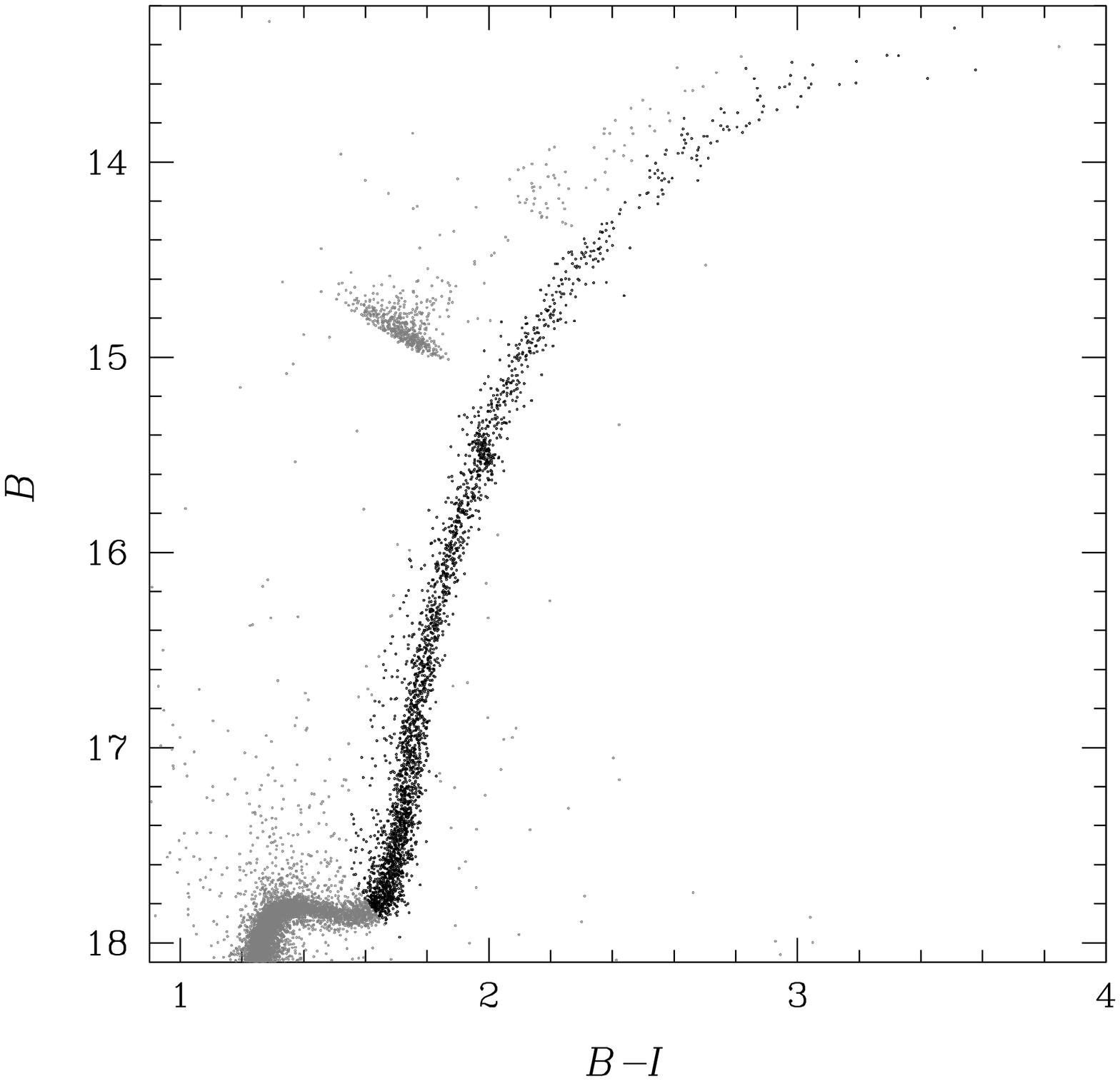}
      \caption{
%
         Similar to what was shown in Fig.~\ref{RGBhst3Ma}, but now
         using the ground-based
         photometry for the RGB stars.  The left panel is a ($B,B-I$)
         CMD in which the black points are the RGB stars that will be
         used for the separation of RGBa and RGBb.  On the right is the
         CMD in $B$ vs.\ $U+I-B$, with a red fiducial line drawn through
         the RGBb stars.}
         \label{3MAG}
   \end{figure*}
%

\begin{figure}[ht!]
   \centering
   \epsscale{0.75}
   \plotone{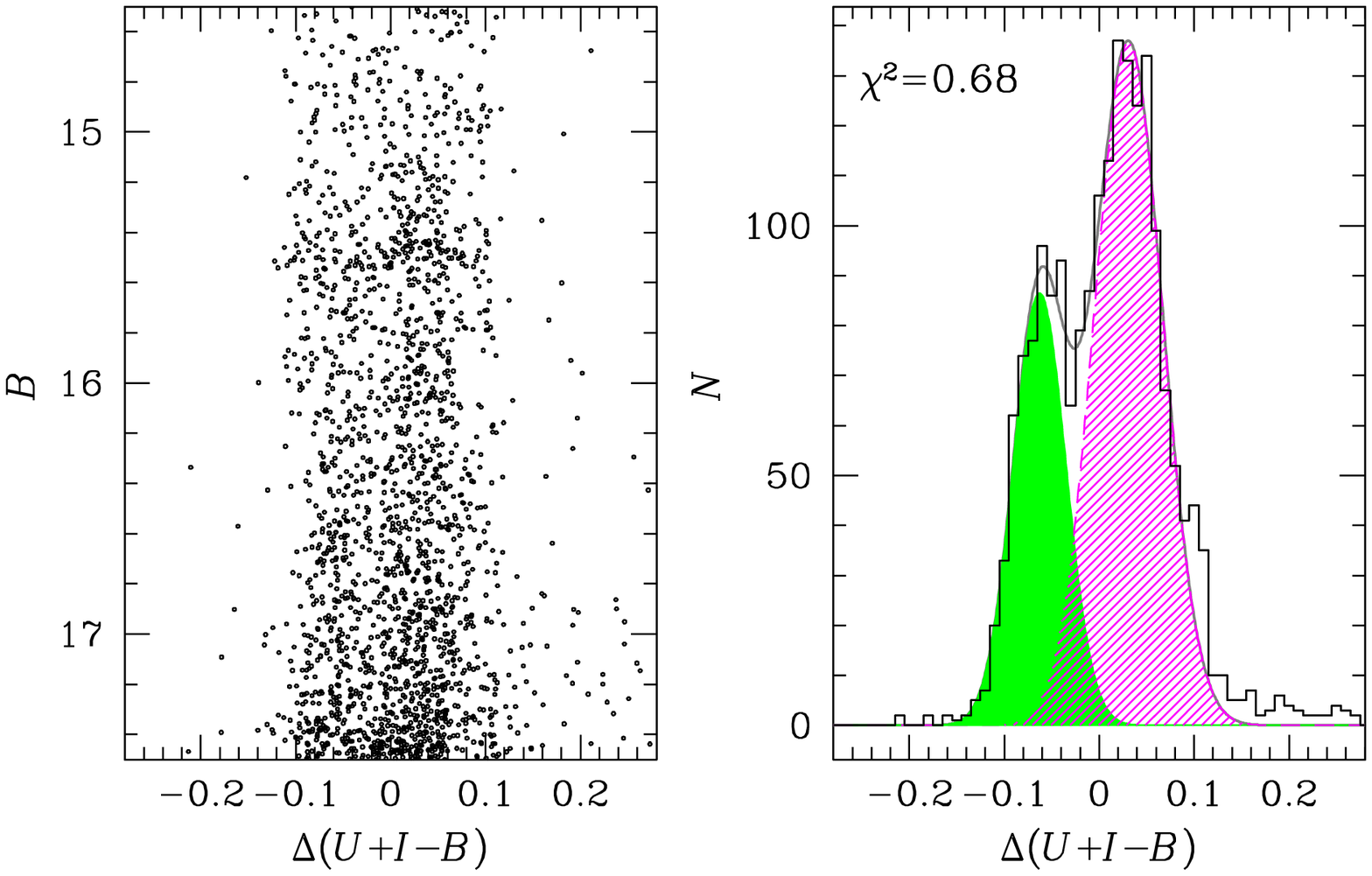}
      \caption{\textit{Left}: same diagram as in Fig.~\ref{3MAG} but
        after subtracting the RGB fiducial. \textit{Right}: histograms
        of $\Delta(U+I-B)$ for RGB stars in our ground-based photometry.}
         \label{HallGB}
   \end{figure}

Finally, in Figure ~\ref{3MAG}, which is analogous to
Figure~\ref{RGBhst2C}, we use $U$, $B$, and $I$ magnitudes to separate
RGBa and RGBb stars in our entire ground-based field, in exactly the
same way as we demonstrated with our {\it HST} photometry using the
corresponding
F336W, F435W, and F814W passbands.  Then we calculate the difference
between the $U+I-B$ value of each star and the value of the fiducial
line at the magnitude of the star, $\Delta( U+I-B)$.  This process makes
the RGB vertical, and we show that in the left panel of
Figure~\ref{HallGB}. The corresponding histograms are plotted in the
right panel of the figure, again fitted by two overlapping Gaussians.

%

\begin{figure}[ht!]
   \centering
   \epsscale{0.75}
   \plotone{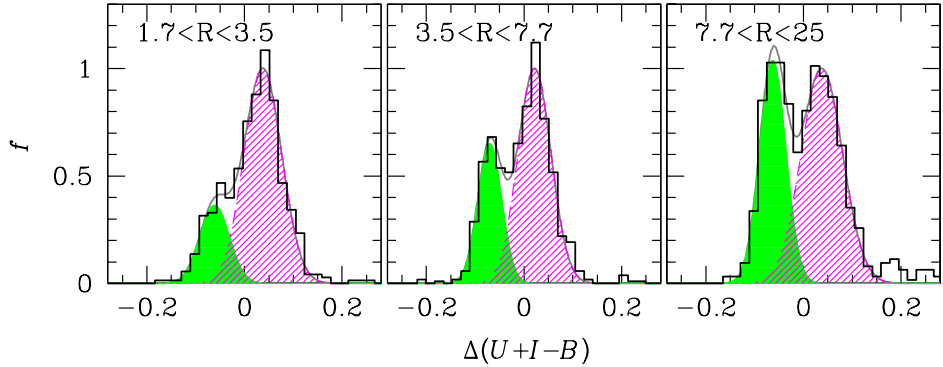}
      \caption{Histogram of the $\Delta$($U+I-B$) distributions
        of RGB stars at three different radial distances.}
         \label{RDRGB}
   \end{figure}

Having developed a method of using ground-based filters for separating
populations, we now study the spatial distribution of
the two RGB components,
by dividing our ground-based field of view into three annuli, each
containing about the same number of RGB stars,
and applying the procedure described above to each group separately. The
histograms of the $\Delta(U+I-B)$ distributions are shown in
Figure ~\ref{RDRGB}.
It is clear from this figure that RGBb is more centrally concentrated
than RGBa, with the fraction of RGBa stars ranging from $0.22\pm 0.04$
in the 1.7--3.5 arcmin bin, to $0.32\pm 0.04$ at radial
distances from 3.5 to 7.8 arcmin, up to $0.43\pm 0.04$ for stars in the
7.8--25 arcmin bin.  This result will be discussed in detail in
Section~\ref{sec:RD}, where we also compare the radial distributions of
the two RGB components with those of MS and HB components.
%

We note finally that we carefully analyzed the SGB and the MS in the $B$
vs.\ $U+I-B$ diagram, the same as we have just done for the RGB, but for
those other regions we found no evidence for multiple sequences,
presumably because of the larger errors in the ground-based photometry
of those fainter stars.

\subsection{The chemical content of stars in the two RGB sequences}

\begin{figure}[ht!]
  \centering
   \epsscale{0.75}
   \plotone{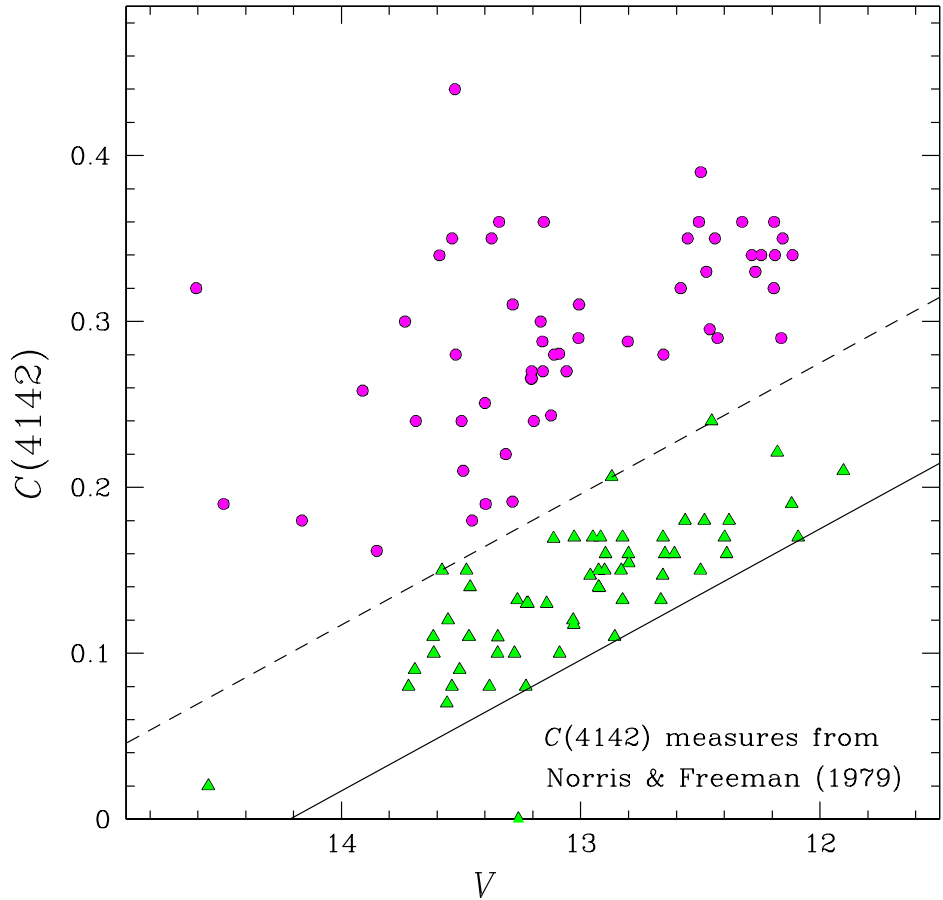}
     \caption{Adaptation of Fig.~1 of Norris \& Freeman (1979), showing
       C(4142) index vs.\ $V$ magnitude for their sample of RGB
       stars. Their full line follows the lower bound of the data,
       while our dashed line separates the CN-weak stars (green
       triangles) from the CN-strong ones (magenta circles).}
        \label{norrisRGB}
  \end{figure}

\begin{figure}[ht!]
   \centering
   \epsscale{0.75}
   \plotone{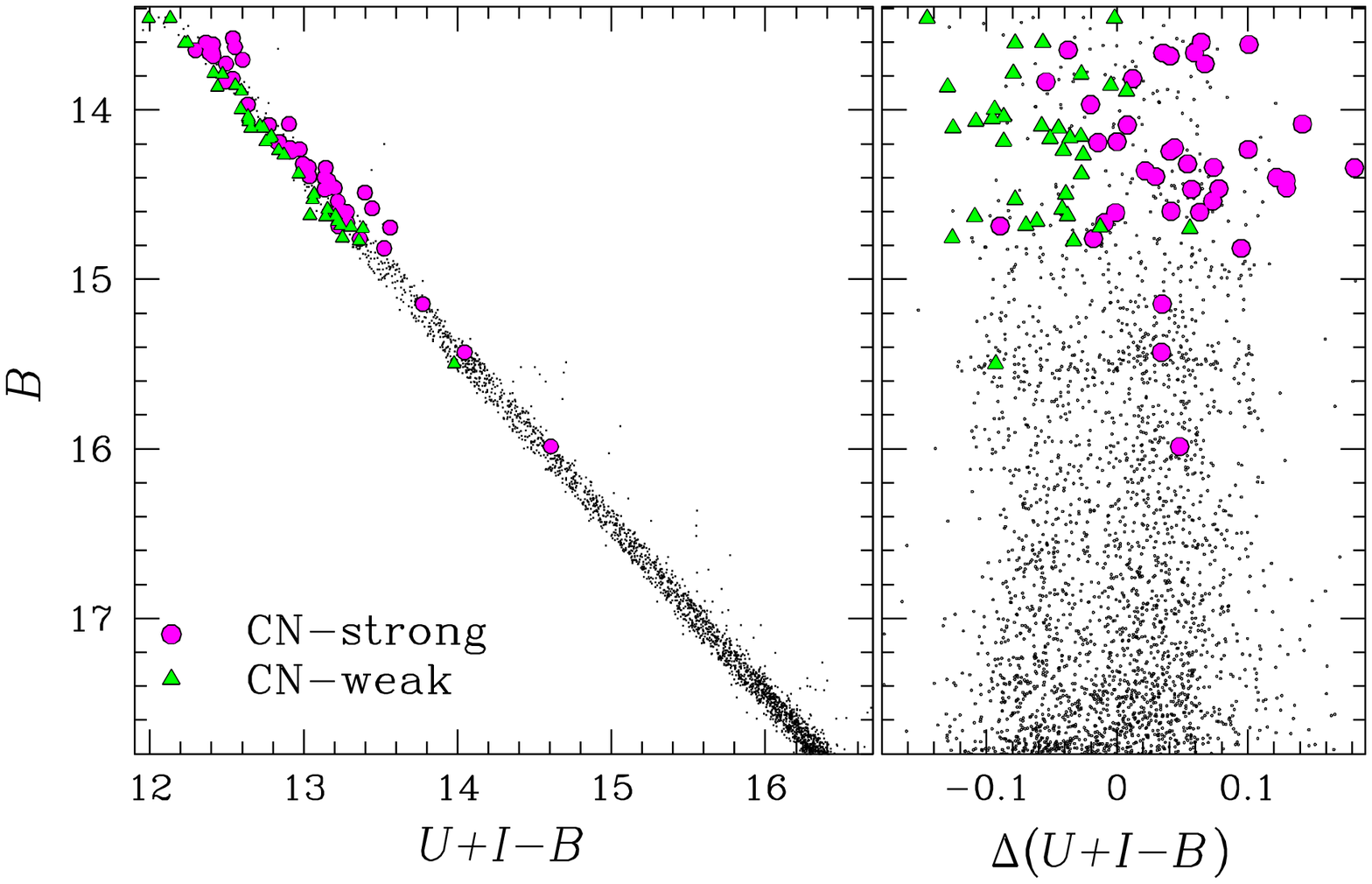}
      \caption{The CN-strong and CN-weak stars defined in
        Fig.~\ref{norrisRGB}, marked in our $B$ vs.\ $U+I-B$ plane (left
        panel) and $B$ vs.\ $\Delta(U+I-B)$ plane (right
        panel).}
         \label{cn3RGB}
   \end{figure}

Since the red giants are the brightest stars in 47 Tuc, their
spectroscopy has a long history.  In the early seventies, large
star-to-star cyanogen variations showed that the cluster is not
chemically homogeneous (e.g., McClure \& Osborn 1974; Bell, Dickens, \&
Gustafsson 1975; Hesser, Hartwick, \& McClure 1977; Norris 1978; Hesser
1978), and in a paper that set a standard of spectroscopic accuracy at
that time, Norris \& Freeman (1979) measured CN in 142 RGB stars of 47
Tuc and found a bimodal distribution, with CN-strong stars more
centrally concentrated.
Our Figure~\ref{norrisRGB} is adapted from their Figure~1, and here we
plot their values of the DDO C(4142) index (which is sensitive to the
blue CN bands) against our $V$ magnitude.  The dashed line separates the
CN-strong and CN-weak stars, which for added emphasis we have indicated
by magenta circles and green triangles, respectively.  We have
80 RGB stars in common with Norris \& Freeman, and
in Figure~\ref{cn3RGB} we have marked their
CN-strong and CN-weak stars with distinctive symbols as defined above, in 
a repeat of our Figure~\ref{3MAG}.
Note that most of the CN-strong stars belong to RGBb, and nearly all the
CN-weak stars lie on RGBa.  It is obviously very tempting to associate
the two groups of CN-weak and CN-strong stars with the two RGBs.  More
accurate measurements would be needed, however, to establish whether the
few exceptions to this rule are real, or are due to errors in either the
photometric or the spectroscopic measures.

%

%

It is now well established that a chemical signature of
multiple stellar populations in GCs is offered by the Na-O anticorrelation
that has been noticed in virtually every cluster observed so far. 
   \begin{figure}[ht!]
   \centering
   \epsscale{0.75}
   \plotone{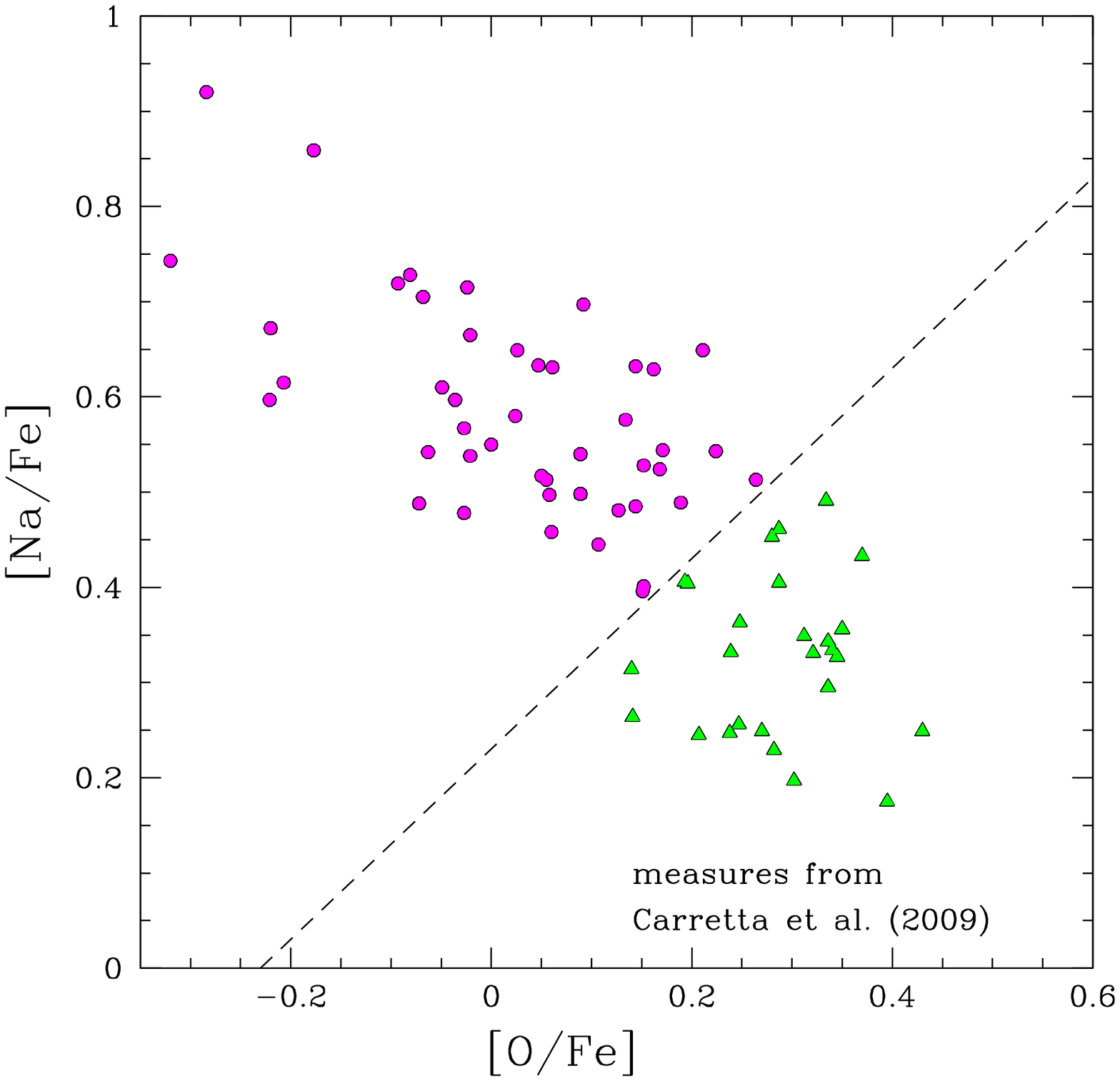}
      \caption{Sodium-oxygen anticorrelation for RGB stars, from
        Carretta et al.\ (2009a). The dashed line arbitrarily separates
        Na-rich/O-poor stars (magenta circles) from Na-poor/O-rich stars
        (green triangles).}
         \label{NaO}
   \end{figure}
%
The Na-O anticorrelation in RGB stars of 47 Tuc has recently been
studied by Carretta et al.\ (2009a), and their data are shown in
Figure~\ref{NaO}, where we have made an arbitrary division into
Na-poor/O-rich stars (green triangles) and Na-rich/O-poor stars (magenta
circles).  The {\it U} vs.\ $ U-B$ CMD shown in Figure~\ref{NaOcmd}
gives photometric evidence for a spread in color that extends from the
base of the RGB to
its tip.  We then identify stars from Carretta et al.\ (2009a) in our
ground-based CMDs of 47 Tuc, and the two groups of stars defined in
Figure~\ref{NaO} are found to segregate on the RGB as illustrated in
Figure~\ref{NaOcmd}, with Na-rich/O-poor stars being systematically
bluer in $U-B$ compared to the Na-poor/O-rich stars, while mixing with
them in the {\it B} vs.\ $B-I$ CMD, in analogy with the results of
Marino et al.\ (2008) for M4.
   \begin{figure*}[ht!]
   \centering
   \epsscale{0.75}
   \plotone{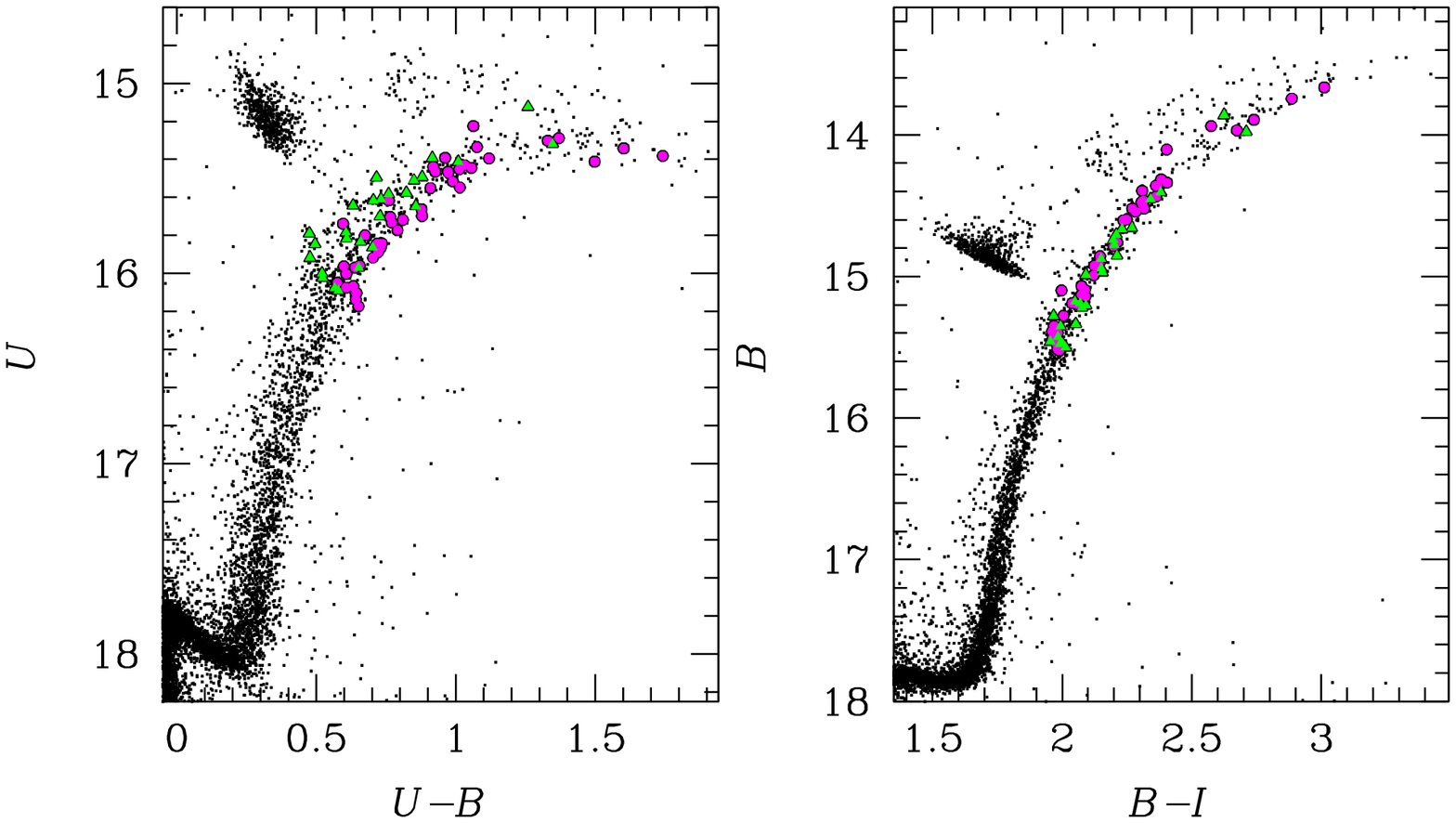}
      \caption{{\it U} vs.\  $U-B$ (left panel) and {\it B}
        vs.\ $B-I$ (right panel) CMD from ground-based photometry.
      The stars belonging to the two groups of Na-rich (O-poor)
      and Na-poor (O-rich) stars are represented  with magenta circles and green
      triangles, respectively.  }
         \label{NaOcmd}
   \end{figure*}
%

\begin{figure*}[ht!]
   \centering
   \epsscale{0.75}
   \plotone{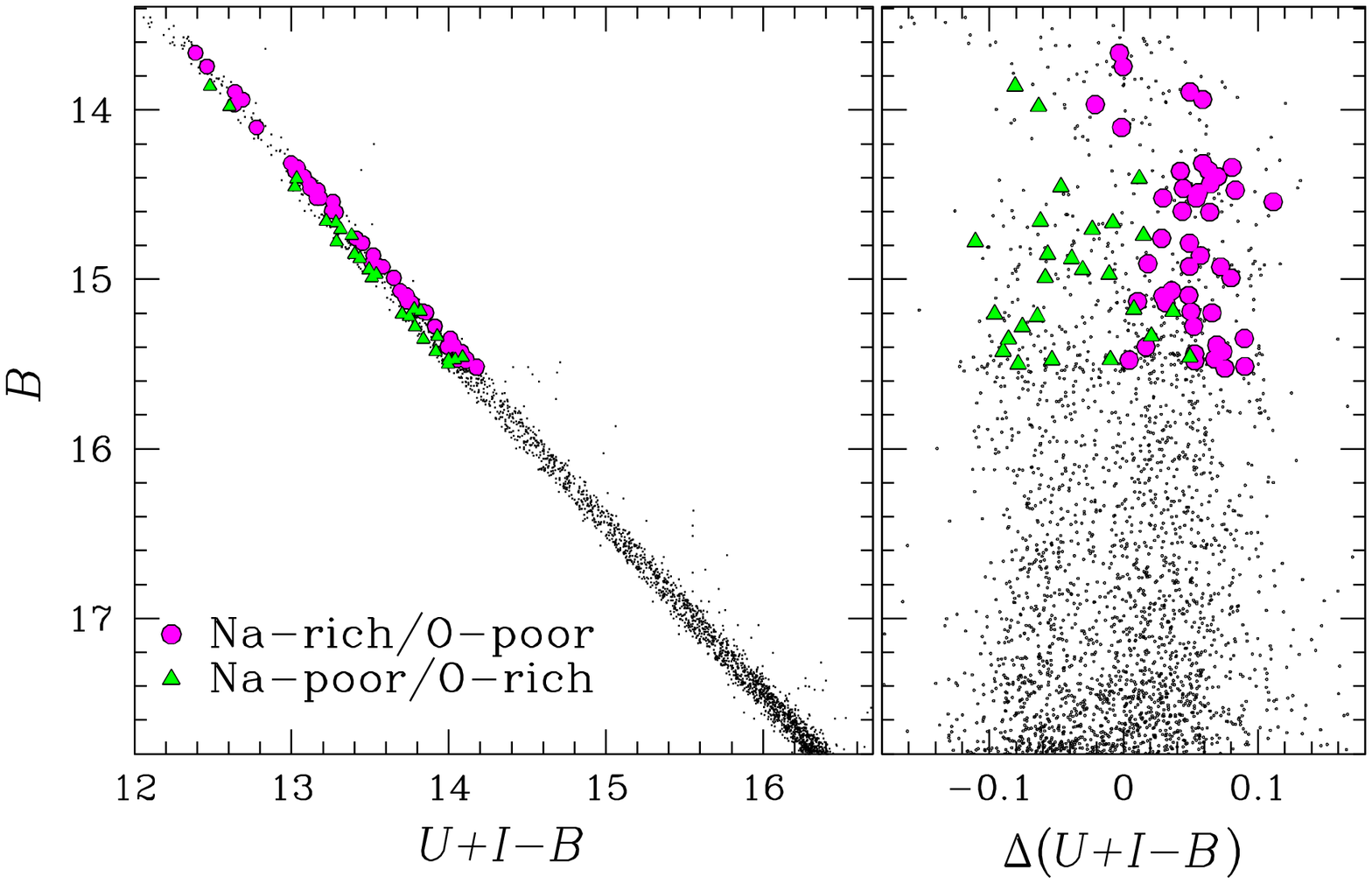}
      \caption{The Na-poor/O-rich and Na-rich/O-poor RGB stars defined
        in Fig.~\ref{NaO} are plotted here in $B$ vs.\ $U+I-B$ (left
        panel); the right-hand panel shows a verticalized version, in
        our usual way.}
         \label{NaO3m}
   \end{figure*}

Finally, in Figure~\ref{NaO3m} we plot the Na-poor/O-rich and
Na-rich/O-poor stars in the {\it B} vs.\ $U+I-B$ diagram and in the {\it
B} vs.\ $\Delta(U+I-B)$ diagram, to investigate the abundances of Na and
O in the two RGBs.  We find that RGBa is populated mainly by
Na-poor/O-rich stars, while all the Na-rich/O-poor stars belong to RGBb.

%

\section{Multiple stellar populations on the Horizontal Branch}
\label{sec:HB}

   \begin{figure*}[ht!]
   \centering
   \epsscale{0.75}
   \plotone{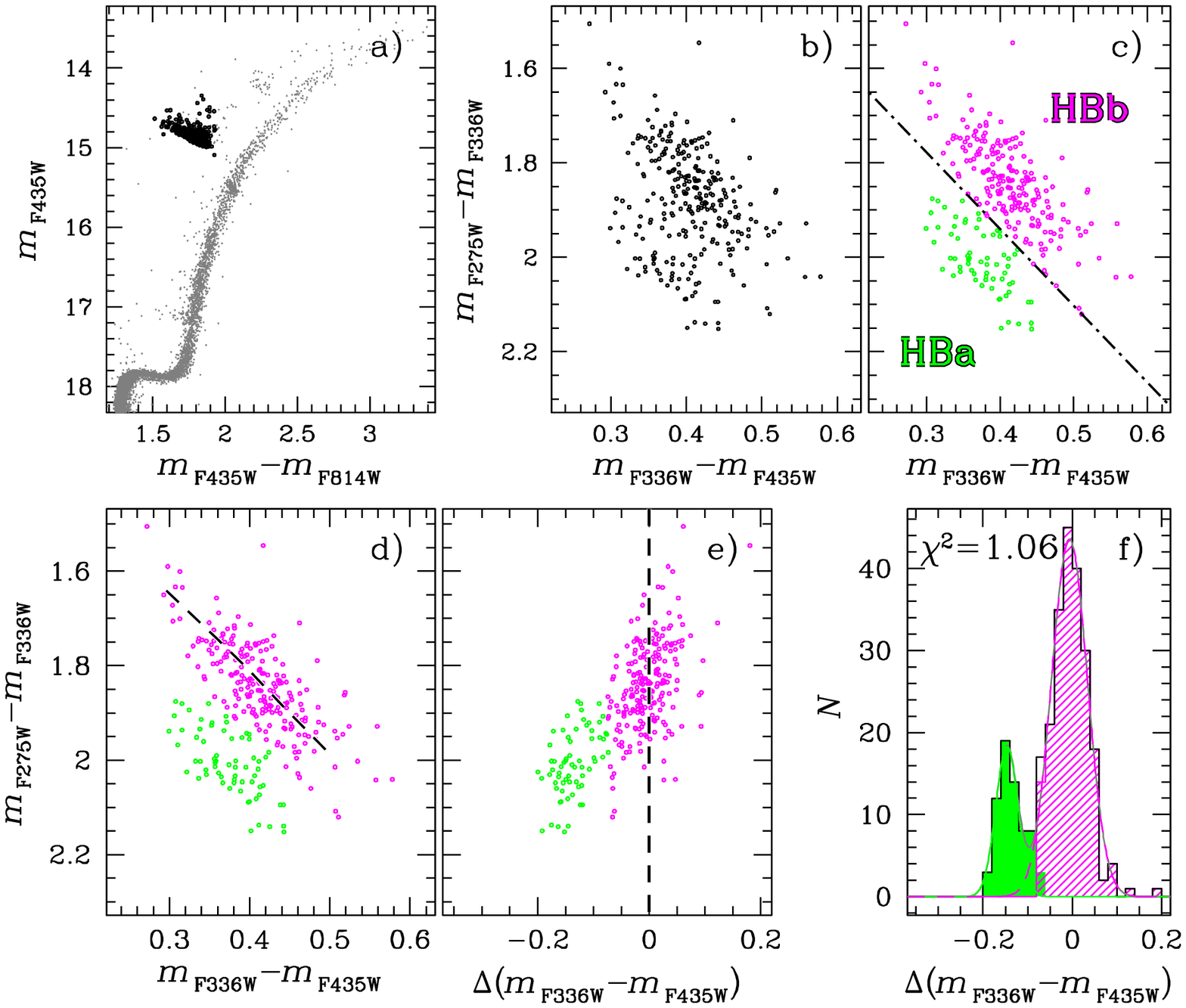}
%
      \caption{
        The HB stars highlighted in black in panel {\it a} can be seen
        to have a bimodal distribution in the two-color diagram of panel
        {\it b}, and in the next panel they are separated into two
        components.  In the bottom row of panels their distribution in
        color is fitted with two Gaussians; see text for details.}
         \label{HBhst}
   \end{figure*}
%

We now turn to the horizontal branch of 47 Tuc. 
Figure \ref{HBhst} shows
the {\it HST} data in the same plots as were used for the other sequences.
In order to point out the stars that we are studying here, we show in
panel {\it a} a long-wavelength CMD, with the HB emphasized in black.
In panel {\it b} we show the ultraviolet two-color diagram of these
stars, with the two HB sequences HBa and HBb
identified as usual in panel {\it c}.  In panel {\it d} we show the
same plot again, but with a fiducial line through the HBb locus.  We
verticalize the sequences in the
usual way in panel {\it e}.  Panel {\it f} shows the corresponding
histogram, with the fit by two Gaussians that are colored appropriately.
Perhaps quite unsurprisingly at this stage, the result that HBa makes
up  $19\pm3$\% of the HB stars is very
similar to what we found for the RGB.

%
   \begin{figure*}[ht!]
   \centering
   \epsscale{0.75}
   \plotone{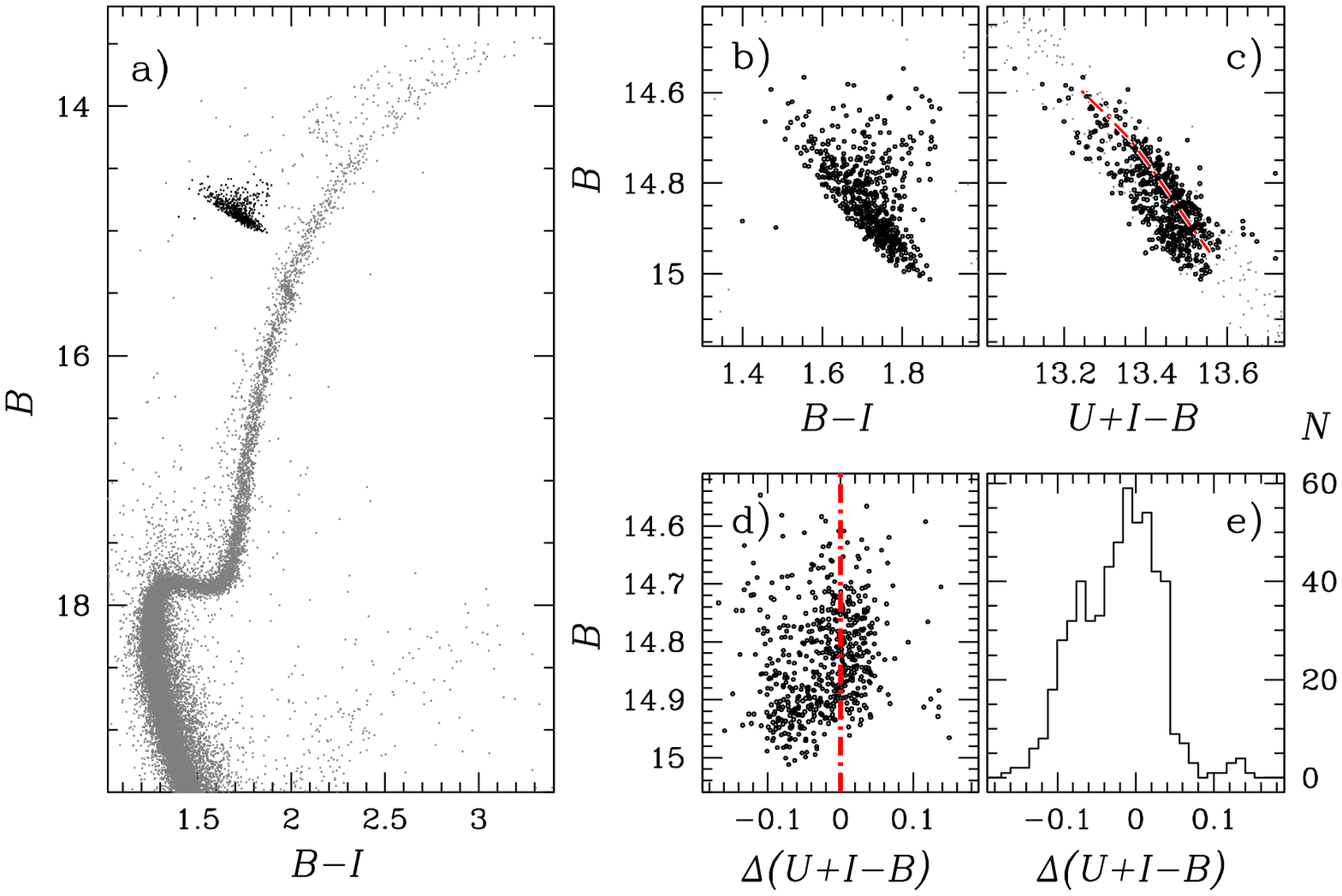}
      \caption{ \textit{Panel a}:\ {\it B} vs.\ $B-I$ CMD from
        ground-based photometry. HB stars are marked in black. A zoom of
        this CMD around the HB region is plotted in panel {\it
          b}. \textit{Panel c}:\ HB stars in the ({\it B}, $U+I-B$)
        plane; the red dashed line is the
        HBb fiducial drawn by hand. \textit{Panels d} and \textit{e}
        show the rectified {\it B} vs.\ $\Delta(U+I-B)$ diagram and
        the histogram of the rectified colors.  }
         \label{HBgb}
   \end{figure*}
%

To study the radial distribution of the two HB populations, we again use
ground-based photometry.  In panel {\it a} of Figure~\ref{HBgb} we show
the $B$ vs.\ $B-I$ CMD of all the stars that passed the
selection criteria described in Section~\ref{data}, and the selected HB
stars are marked in black.  Panel {\it b} shows a zoom of the HB, in the
same color system, and in panel {\it c} we plot the same stars in the $B$
vs.\ $U+I-B$ diagram, where the distribution shows some bimodality.  
The red dashed line is the fiducial line of the more populous HB
component,
and by analogy with the previous figure we refer to the lower-left stars
as HBa and the upper-right stars as HBb.  The rectified $B$ vs. $U+I-B$
plot is shown in panel {\it d}, and the corresponding histogram is shown
in panel {\it e}.  

   \begin{figure}[ht!]
   \centering
   \epsscale{0.75}
   \plotone{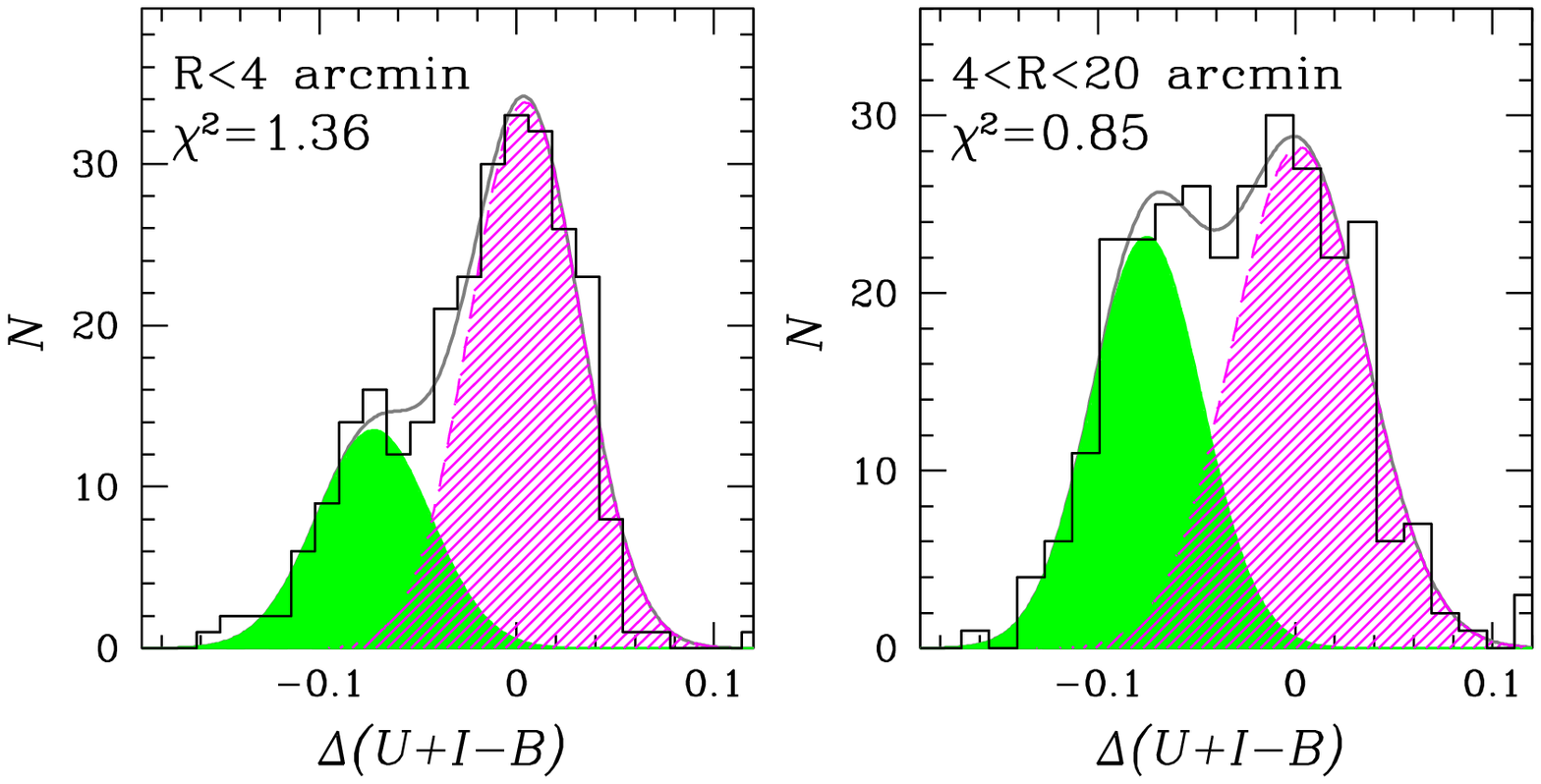}
      \caption{ Histogram of the $\Delta$($U+I-B$) distribution of HB
      stars from ground-based photometry at two radial distances from
      the cluster center.}
         \label{RDHB}
   \end{figure}
%

It has been suspected for a long time that the HB of 47 Tuc might
contain multiple populations.  Norris \& Freeman (1982) measured the
strengths of CN and CH bands in the spectra of 14 HB stars and concluded
that their results were ``consistent with a dichotomy as found for the
giants'', which agrees with the result that we have just presented.
They also noted that CN-weak stars are on average $\sim$ 0.04 mag
brighter in $V$ than CN-strong stars --- which is similar to what we see
in panel {\it e} of Figure~\ref{HBhst} and panel $d$ of
Figure~\ref{HBgb}.
%

   \begin{figure}[ht!]
   \centering
   \epsscale{0.75}
   \plotone{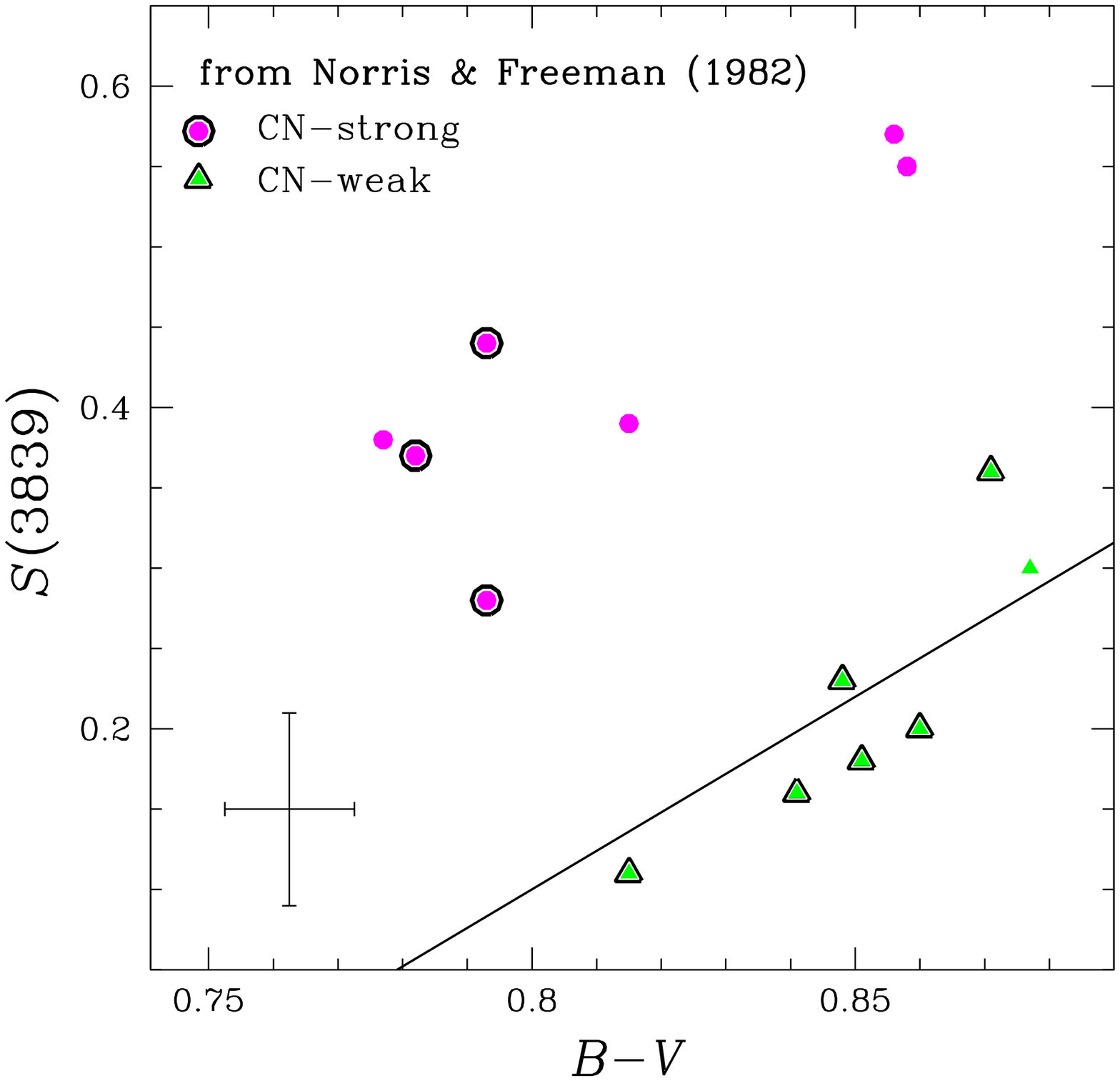}
      \caption{S(3839) index for 14 HB stars as a function of {\it B-V} from
        Norris \& Freeman (1982). The line shows the dependence of
        S(3839) on the effective temperature as proposed by
        these authors. The CN-weak and CN-strong stars are
        plotted with green triangles and magenta circles, respectively, while
      we have marked with black 
black contours
the nine stars for which  we have {\it U},
      {\it B} and {\it V} photometry.
}  
         \label{CNNF82}
   \end{figure}
%
Fig.~1a from Norris \& Freeman (1982) is reproduced here as
Fig.~\ref{CNNF82}. We plotted their S(3839) index measurements for 14 HB
stars as a function of the {\it(B-V)} color and drew the line that
shows 
the dependence of S(3839) on 
effective temperature, as
discussed by Norris \& Freeman (1982).  The two groups of CN-weak and
CN-strong stars defined by Norris \& Freeman are plotted as green
triangles and magenta circles, respectively.  Our ground-based catalog
has in common nine out these fourteen stars.  These stars are marked
with black
circles.
Among them three are CN-strong and
six CN-weak stars.

\begin{figure}[ht!]
   \centering
   \epsscale{0.75}
   \plotone{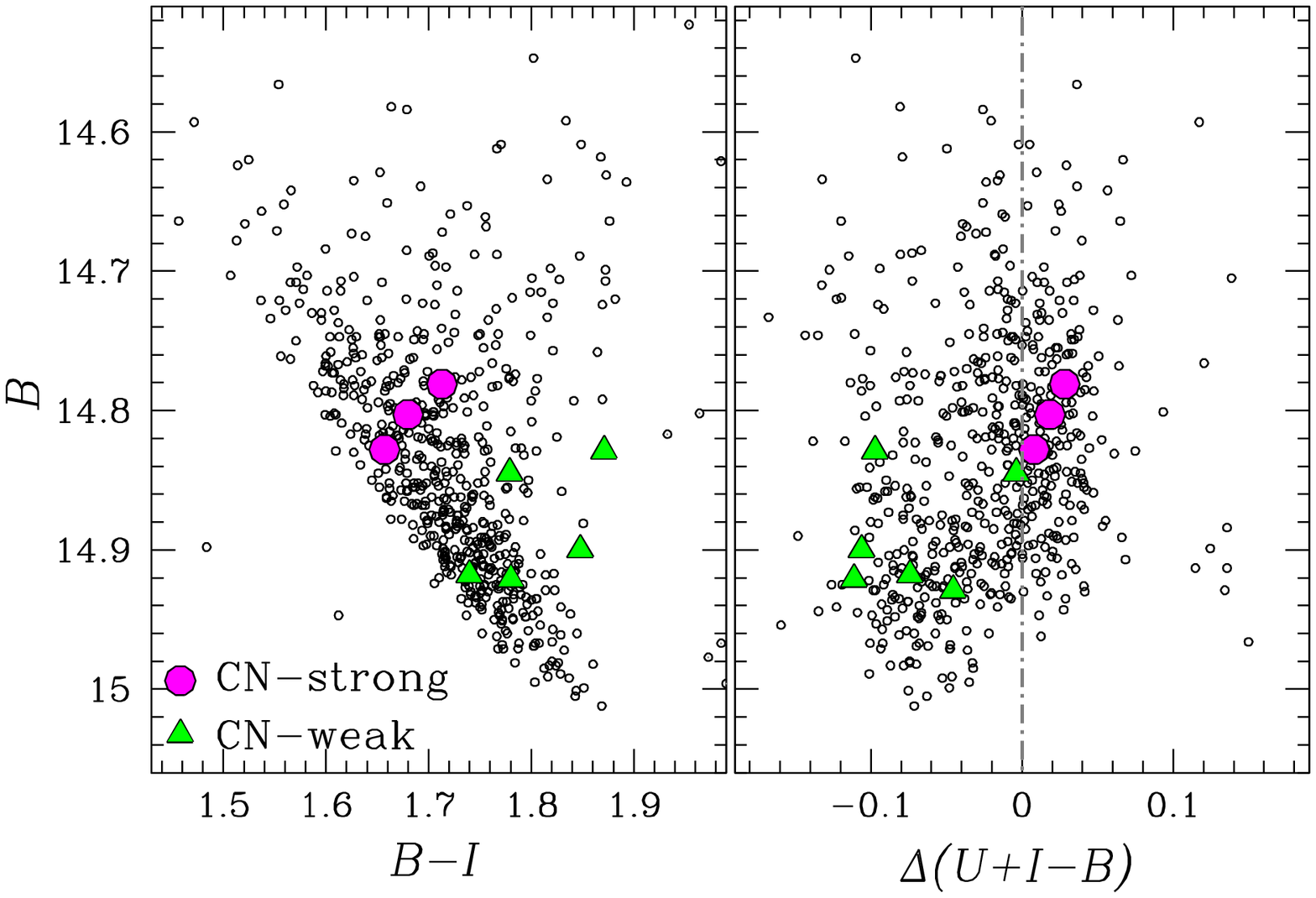}
      \caption{Reproduction of the {\it B} versus {\it B-I} CMD (left
        panel) and the {\it B} versus $\Delta$({\it U+I-B}) diagram
        (right panel) for HB stars of Figs.~\ref{HBgb}b and
        ~\ref{HBgb}d. The two groups of CN-weak and CN-strong stars
        as defined in Fig.~\ref{CNNF82} are plotted with blue triangles
        and red circles.}
         \label{HBCN}
   \end{figure}

In Fig.~\ref{HBCN} we show the position in the {\it B} vs. {\it B-I} 
CMD (right panel) and in the {\it B} vs.\ {$\Delta(U+I-B)$} diagram 
the nine stars in common (left panel). CN-strong stars are on average 
 brighter than CN-weak ones in the {\it B} band. The right panel
shows that CN-strong and CN-weak stars have {$\Delta(U+I-B)$} values
consistent to that of the bulk of the HBb and HBa groups, respectively.
An exception to this rule is given by a CN-weak star 
which has a {\it B} magnitude and {$\Delta(U+I-B)$} value close to that of
 HBb stars. It is not clear if this anomalous position in the latter
 diagram is an intrinsic property of this star or is due to either
 photometric or spectroscopic errors.
%
%

The connection between the HB morphology in GCs and the groups
of stars with different abundances of light elements (Na, O, C, N)
is not a peculiarity of 47 Tuc but has been also observed in the GC
NGC 6121 (M4).
As already mentioned in Sect.~\ref{introduction}, this cluster hosts
two groups of stars with different Na and O content (Marino et
al.\ 2008), as well as a bimodal HB.
Marino et al.\ (2010) have recently measured oxygen and sodium
for stars in the blue and the red HB segments of M4,
and found that the red HB is made by stars with low
Na and high O content, while blue HB stars are all
Na-enhanced and O-depleted (possibly He-rich).   

These results on 47 Tuc and NGC 6121 provide direct evidence that the HB
morphology of these GCs is strictly related to the multiple stellar
generations they host, and suggest that the multiple sequences
discovered in the CMDs of many GCs may be connected with the HB
morphology.

\subsection{The role of C, N, and O on the SGB and the HB}
\label{CNOsgbhb}

In order to better understand the origin of the multicolor distribution
of SGB and HB stars, we present here an analysis similar to that
described in Section~\ref{YcnoMS} for MS stars, using a number of
synthetic spectra over the wavelength range of interest 
(i.\ e., $2000 < \lambda < 5000$~\AA).
They were computed using the Kurucz (1993)
model atmospheres (with the overshooting option switched off), and line
lists from the Kurucz CD-ROMs. Although these lists may be incomplete,
especially in the UV, and the models adopt the 1-dimensional
approximation, the resulting model atmospheres are still useful for the
present purposes, which consist of identifying the major spectroscopic
features that can affect the photometry of stars in 47 Tuc.  In the rest
of this discussion, we limit ourselves to consideration of differential
effects between stars that should have very similar spectra
unless their chemical compositions differ: a subgiant star with 
$T_{\rm eff}=5700$~K and $\log{g}=3.75$, and a red HB star with 
$T_{\rm eff}=5400$~K and $\log{g}=2.60$.

%
   \begin{figure}[ht!]
   \centering
   \epsscale{0.75}
   \plotone{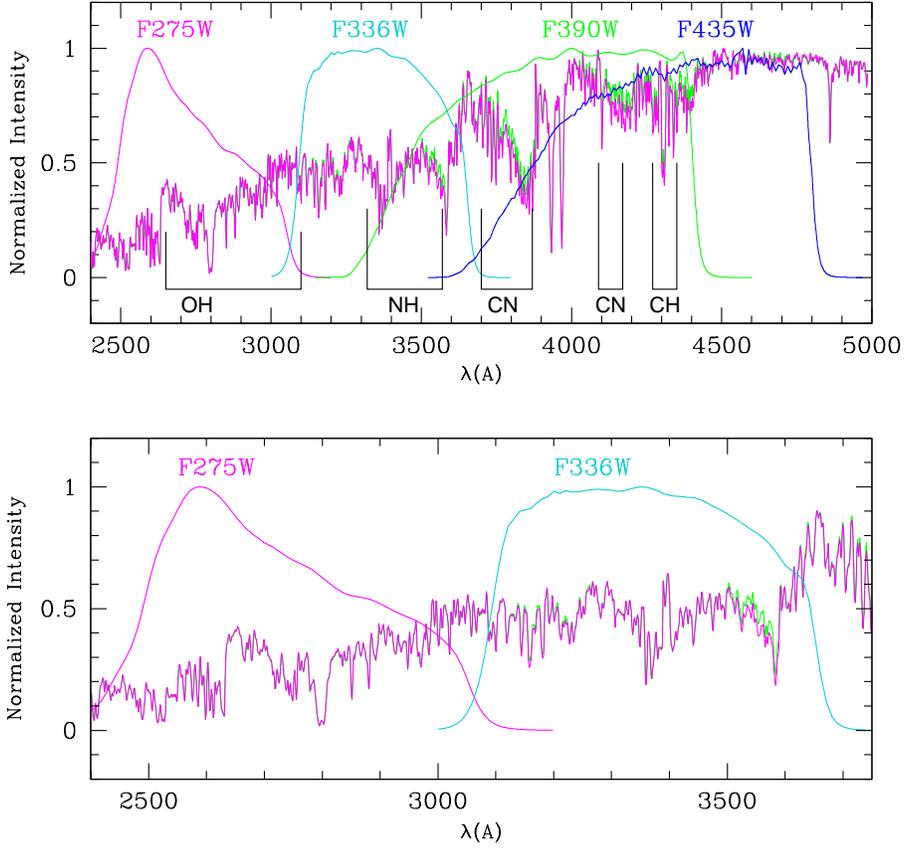}
      \caption{\textit{Top Panel}: Comparison between two synthetic
        spectra: one for a N-rich star (magenta) and one for a N-poor
        star (black). The spectra are given as flux (in arbitrary units)
        and are smoothed at 1\AA\ resolution for clarity; they have been
        computed for parameters typical of a subgiant star in 47 Tuc,
        with chemical compositions given in the text. For reference, the
        normalized throughputs of the bluest broad-band filters of
        WFC3/UVIS F(275/336/390/435/475)W are also shown. Labels on the
        bottom indicate the wavelength range where important
        spectroscopic features involving CNO elements cause significant
        absorption. The most important contributions come from OH at
        $\sim$2600-3100\AA\ and NH at $\sim$3300-3600\AA. \textit{Bottom
        Panel}: a zoom-in of the spectral region that is of particular
        interest for the present paper.}
         \label{f:band}
   \end{figure}
%

One set of synthetic spectra was computed with the typical abundance
pattern observed in metal-poor stars in the field, [C/Fe] = [N/Fe] =
0.0, [O/Fe] = 0.4; spectra computed with these
parameters are called ``N-poor''. 
They should mimic the composition of the stars that Carretta et al.\
(2009a) call primordial-generation.
We computed another pair of spectra for the same atmospheric parameters,
now with [C/Fe] = $-$0.2, [N/Fe] = +1.3, and [O/Fe] = $-$0.1, but with
the other element abundances the same as in the ``N-poor'' stars. We
call this second group ``N-rich''; they should mimic the spectra of the
stars that Carretta et al.\ (2009a) call second-generation.  
 For both the ``N-poor'' and ``N-rich'' stars we assumed 
[Fe/H] = $-$0.75 and [$\alpha$/Fe] = 0.4.
The spectra were then integrated over the transmission of {\it HST} 
filters F275W, F336W, F390W, and F435W, to derive the fluxes expected 
in those bands.

Figure~\ref{f:band} compares a pair of synthetic spectra (those
corresponding to subgiant stars), and shows the transmissions of the
filters. This figure shows that the differences between the ``N-poor''
and ``N-rich'' spectra are essentially due to different strengths of the
molecular bands: The OH band (in the wavelength range 2600--3200 \AA),
is stronger in ``N-poor'' stars and falls within the F275W band; the NH
band at $\sim$ 3400 \AA\ is stronger in the ``N-rich'' spectra and falls
within the F336W band.  The CN violet bands (stronger in N-rich spectra)
at 3883~\AA\ and 4216~\AA, and the CH G-band (stronger in N-poor
spectra) falls in the F390W band. The last two molecular bands are also
within the F435W pass-band.  As a consequence, the flux predicted for
both the F336W and the F390W bands is smaller for the ``N-rich'' spectra
than for the ``N-poor'' ones. The difference is greater for the F336W
band, where it
can
be as much as 0.1 mag. 
The opposite holds for the F275W band.  Abundance variations of 
C, N, O elements do not appreciably affect the stellar flux for 
passbands at longer wavelengths. 

We can now compare our theoretical values of three UV colors with the
observed ones:
\begin{itemize}
\item $m_{\rm F275W}-m_{\rm F336W}$: This color index is predicted to be
  larger (redder) for N-poor and smaller (bluer) for N-rich stars; this
  is the combined effect of less OH absorption in the F275W band and
  more NH absorption in the F336W band in N-rich compared to N-poor
  stars.  If N varies while He and Mg do not, the predicted color
  difference for the abundances given above is
  0.17 mag for the subgiant, and 0.19 mag for the red HB star.
\item $m_{\rm F336W}-m_{\rm F435W}$: This index is larger (redder) for
  the N-rich and smaller (bluer) for the N-poor stars, again a result of
  stronger NH absorption in the F336W band.  If there is again no
  difference in He and Mg, the predicted color difference is
  0.09 mag for the subgiant and 0.10 mag for the red HB star.
\item $m_{\rm F390W}-m_{\rm F435W}$: This index is larger (redder) for
  the N-rich and smaller (bluer) for the N-poor star, again due to
  stronger NH and especially CN absorption in the F390W band, and less
  CH absorption in the F435W band.  If there is no difference in He and
  Mg, the predicted color difference is
  0.05 mag for the subgiant and 0.07 mag for the red HB star.
\end{itemize}
These predicted differences are indeed similar to those observed in the
two-color diagram in panels {\it b} and {\it c} of Fig.~\ref{HBhst}.

It is of course a complication that additional differences are expected
if the abundances of He and/or Mg
are also different.  The F275W
band includes the very strong resonance doublet of Mg II, so that the
impact of Mg is appreciable (up to $\sim$ 0.05~mag). The main effect of a
helium difference is a change in the effective temperatures of the
stars:\ N-rich stars that are also He-rich are expected to be warmer and
bluer.  Even a small temperature difference has a quite dramatic effect
on the UV bands: a difference of 100~K (corresponding to a change of
$\sim$ 0.07 in $Y$) makes $m_{\rm F275W}-m_{\rm F336W}$ bluer by a
further $\sim$ 0.2 mag, and more than offsets the difference in $m_{\rm
F336W}-m_{\rm F435W}$ color between N-rich and N-poor stars.  Such a
large difference is clearly excluded by the photometry of 47 Tuc, which
on the whole agrees quite well with
only a minimal variation in helium ($\Delta Y \sim$ 0.015), as suggested
in Sect.~\ref{YcnoMS} in our attempt to explain the multicolor
observations of the double MS.

\section{The radial distribution of stellar populations} 
\label{sec:RD}

In the previous sections, we examined the radial gradients  of
     the populations one sequence at a time.  Here, we put all
     the information together to develop a comprehensive picture
     of the cluster.  Since from an abundance perspective, most
     studies have focused on CN, we will frame the discussion in
     terms of that molecule.

The spatial distribution of stellar populations with different CN in 47
Tuc has been widely studied and debated in the literature, and little
doubt remains concerning the presence of a significant radial gradient.
On the basis of CN measurements of 142 RGB stars Norris \& Freeman (1979)
found that the CN-strong population is more centrally concentrated.  But
the same data were further analyzed by Hartwick \& McClure (1980), who
found no evidence of differences in radial distribution of stars with
different CN strength.  Norris \& Smith (1981) 
found CN-strong stars in the majority
in the inner $\sim$ 3 arcmin, a nearly equal fraction of
CN-strong and CN-weak stars between $\sim$ 3 and $\sim$ 15 arcmin, and a
predominance of CN-weak stars at larger radial distances. Langer et
al.\ (1989) concluded that because of the small sample size the results
of Norris \& Freeman (1979) should be considered inconclusive.  Finally,
Briley (1997) studied the radial distribution of the CN-strong and
CN-weak populations on the basis of $\sim$ 300 RGB stars with radial
distances larger than 4 arcmin.  He found that the relative numbers of
CN-rich and CN-poor stars are roughly constant within $\sim$ 13 arcmin of
the cluster center, while the distribution exterior to 13 arcmin is
clearly biased toward CN-poor stars, thus confirming the radial gradient
first detected by Norris \& Freeman (1979).

   \begin{figure}[ht!]
   \centering
   \epsscale{0.75}
   \plotone{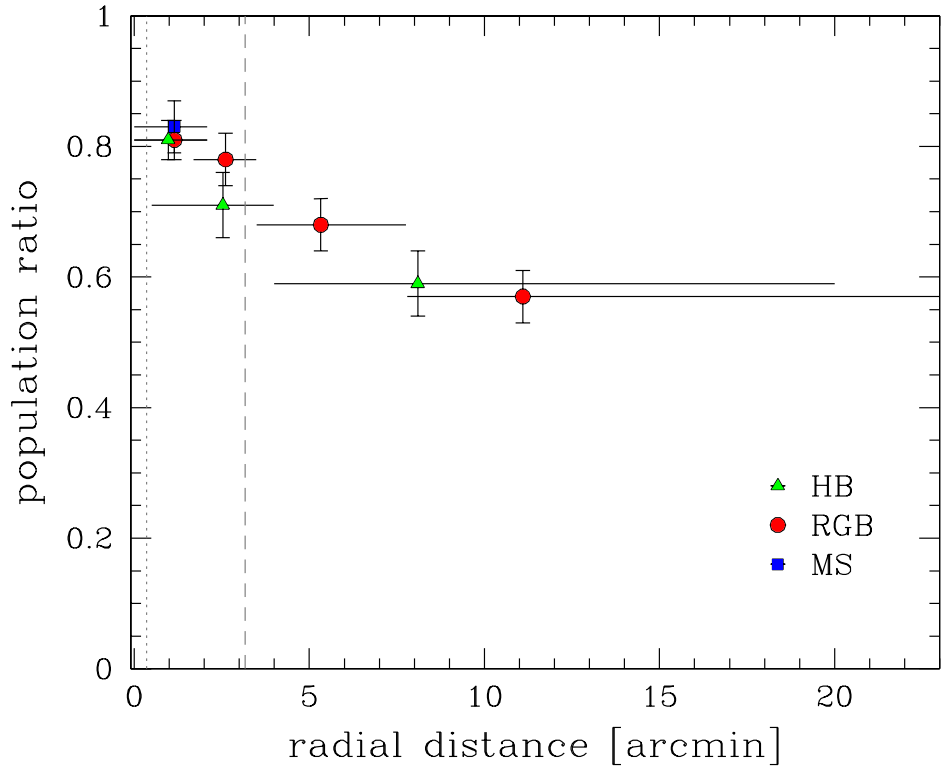}
      \caption{Radial distribution of the fraction of HBb (green
        triangles), RGBb (red circles), and MSb stars (blue square)
        with respect to the total number (component a + component b) of
        HB, RGB, and MS stars, respectively.
The horizontal lines indicate the radial extent of the region
        corresponding to each measure. Vertical dotted and dashed lines
        mark the core and the half-mass radius, respectively.}
         \label{RDall}
   \end{figure}
%

An obvious advantage of photometric measurements is that we can
study more stars and therefore get better statistics.  In the following
we take advantage of the large size of our photometric catalogs to
analyze the radial distributions of the multiple stellar sequences along
the RGB and the HB.  In Section~\ref{sec:RGB} we used both {\it HST} and
ground-based photometry to estimate the RGB population ratio in four
radial intervals, while in Section~\ref{sec:HB} we determined the
fraction of stars in the two HB segments at three radial distances.

The results for the radial distributions of HB and RGB stars are
summarized in Figure~\ref{RDall}, where we have plotted the fraction of
HBb with respect to the total HB stars (green triangles) and the
fraction of RGBb stars with respect to the total number of RGB stars
(red dots).  For completeness we also show, (as a blue square) the
fraction of MSb stars with respect to the total number of MS stars.
(The MS split could be measured only at the center, where we have deep
HST images; the ground-based images are too shallow and crowded to allow
the MS populations to be discerned).  We find that in the central field
the fraction of MSb, RGBb, and HBb stars with respect to the total
number of MS, RGB, or HB stars, respectively, is about 80--82\% for
each.  For the RGB and HB this fraction falls to about 60\% in the outer
parts of the cluster.  Thus the RGBb and HBb populations, which likely
represent a second generation, appear to be more centrally
concentrated. An integration of the $a/b$ population ratio, adopting
a King model appropriate for 47 Tuc, 
reveals that globally the {\it first generation} (MSa, SGBa, RGBa,
HBa) accounts for $\sim 30\%$ of the present-day stellar content of the
cluster, while the {\it second generation} (MSb, SGBb, RGBb, HBb)
accounts for the $\sim 70\%$ majority share of the cluster,
in agreement with the fraction of first and second generation stars
measured by Carretta et al.\ (2009a) on the basis of the Na and O
content.
One final note:\ it should be remembered that the proportions that we
have quoted refer only to the sum of the components that we have called
$a$ and $b$, which together make up only 92\% of the total, the other
8\% being the third population, which we glimpse only as the faint
component of the SGB.  In the other regions of the CMD the third
population presumably makes up some small fraction of the parts that we
call $a$ and $b$.

Such a global predominance of the second generation sets strong
constraints on scenarios for the formation of multiple populations in
GCs, and actually for the formation of GCs {\it tout court}.  The fact
that the second generation is much more centrally concentrated 
has also been observed in other GCs (such as $\omega$ Centauri, Sollima
et al.\ 2007, Bellini et al.\ 2009) and suggests
that much of the first generation (whatever it was) might have been
tidally stripped from the progenitor of 47 Tuc. This is consistent with
a similar suggestion for $\omega$ Cen by Bekki \& Norris (2006).
We note that hydrodynamic plus $N$-body
simulations of the formation of multiple stellar populations in GCs
predict a larger concentration of second generation stars in the cluster
central regions (D'Ercole et al.\ 2008, Decressin et al.\ 2008).

\section{Connecting the multiple sequences along the MS, SGB, RGB, and
  HB.}
\label{sec:con}

So far we have analyzed each evolutionary phase separately from the
others, using the $m_{\rm F275W}-{\it m}_{\rm F336W}$ vs.\ $m_{\rm
F336W}-m_{\rm F435W}$ two-color diagram to separate multiple stellar
populations along each of the evolutionary phases:\ MS, SGB, RGB, and
HB.  The various CMDs show that
in our central {\it HST} field
the stellar distribution is bimodal for all such evolutionary phases, with
about 20\% of the stars of each phase
falling
in the lower left part of the two-color diagram, and the remaining 80\%
populating the upper right part of the diagram.
%
   \begin{figure*}[ht!]
   \centering
   \epsscale{0.75}
   \plotone{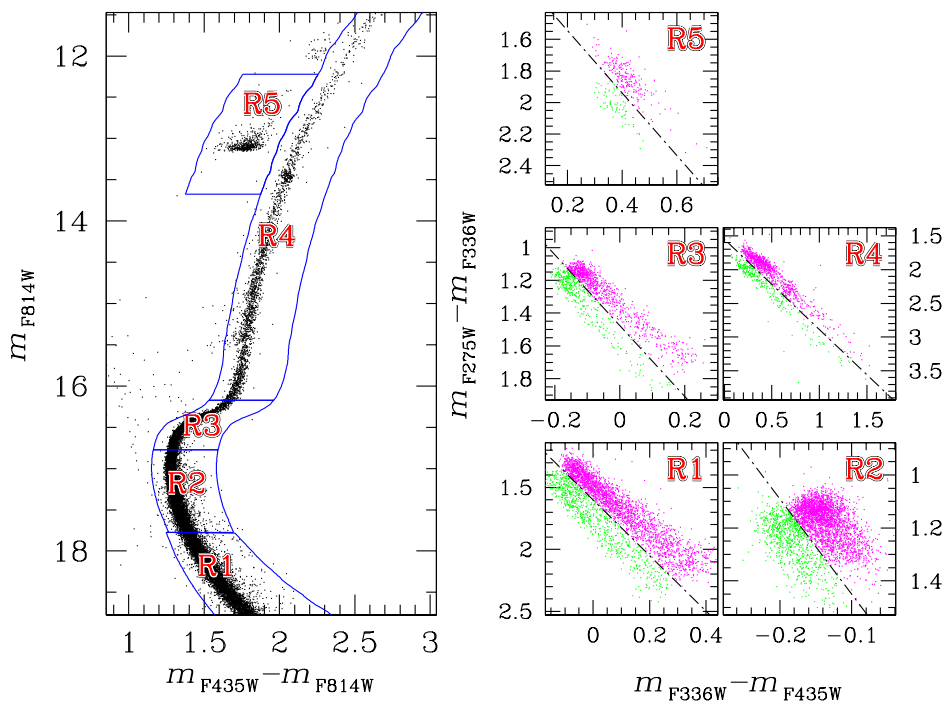}
      \caption{\textit{Left}: $m_{\rm F814W}$ vs.\ ${\it m}_{\rm
          F435W}-m_{\rm F814W}$ CMD from {\it HST} photometry, used to
          define the five regions labeled R1,R2,...,R5. \textit{Right}:
          $m_{\rm F275W}-m_{\rm F336W}$ vs.\ $m_{\rm F336W}-m_{\rm
          F435W}$ two-color diagrams for stars in the five CMD regions.
          Dash-dot lines are used to arbitrarily separate the two
          sequences that are present in each part of the CMD.  }
         \label{allsel}
   \end{figure*}
%

This behavior is summarized in Fig.~\ref{allsel}, where we have used the
CMD in the left panel to select five regions labeled R1--R5,
respectively, marking the MS, the turn-off region, the SGB, the RGB, and
the HB.  In the right panels are plotted the two-color diagrams for
stars in each of these CMD regions, with the black dash-dot lines drawn
so as to separate the two groups of stars that are colored green and
magenta, just as in all previous Sections.

\begin{figure*}[ht!]
   \centering
   \epsscale{0.45}
   \plotone{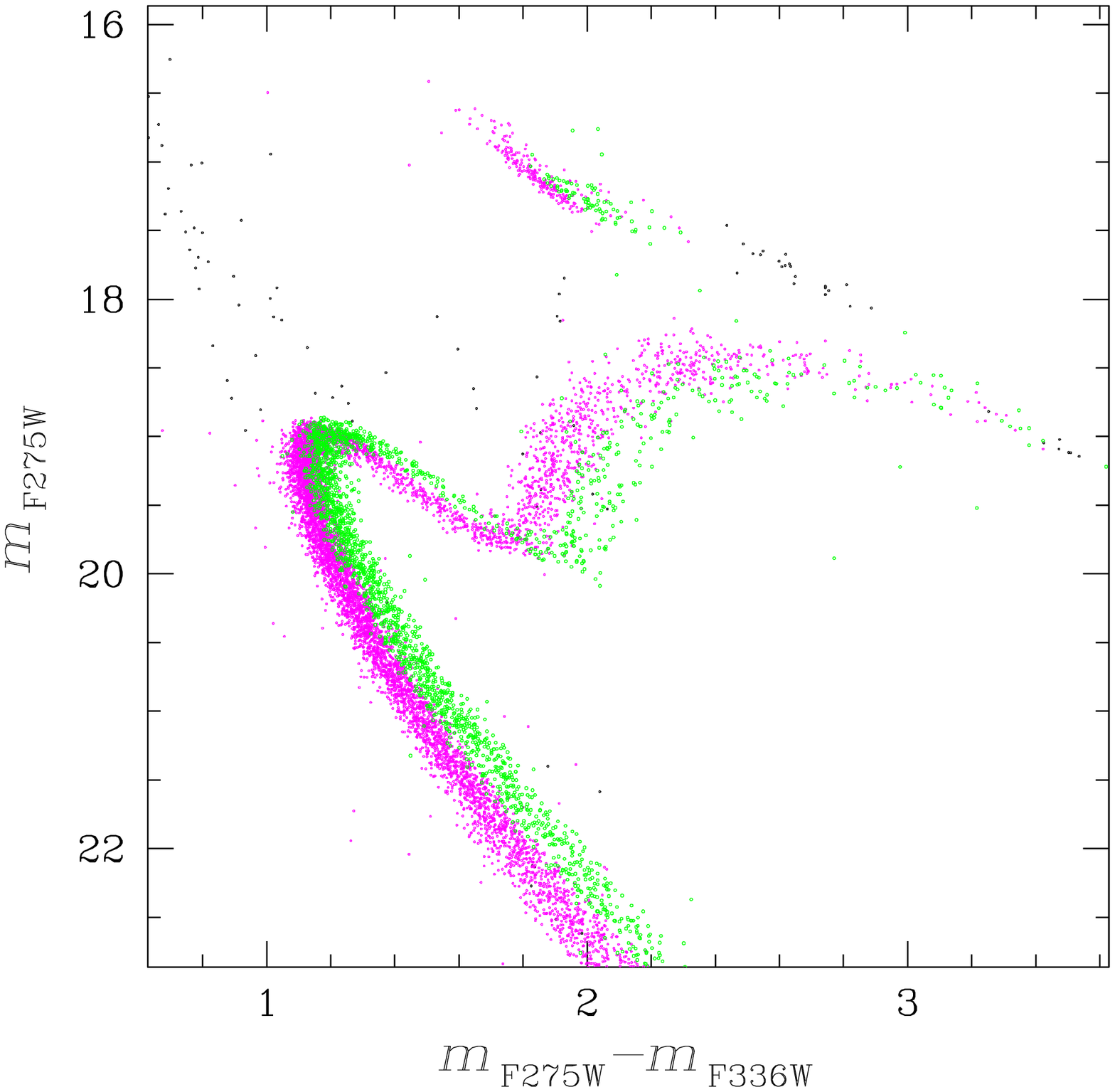}
   \plotone{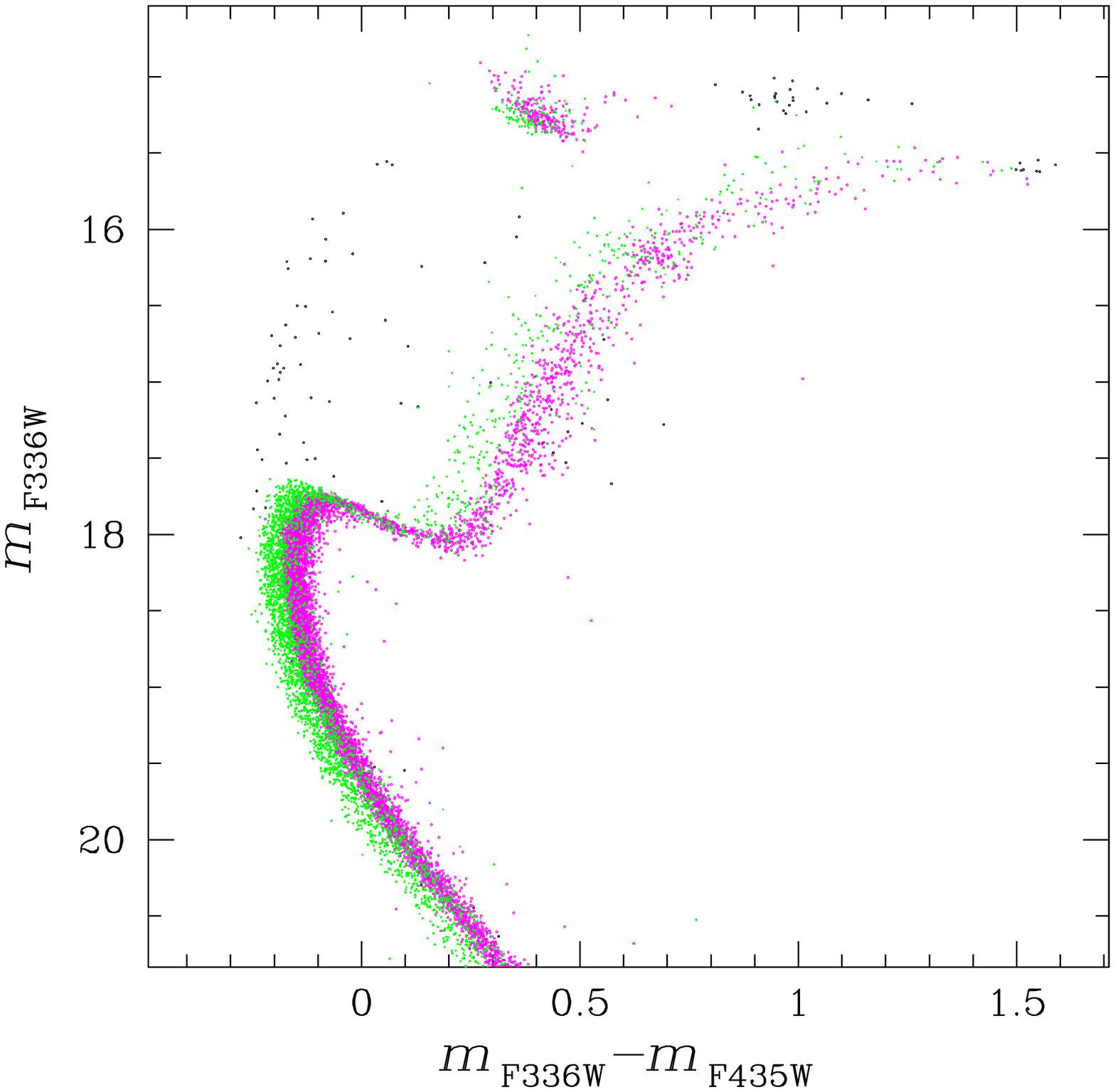}
      \caption{ CMDs with $m_{\rm F275W}$ vs.\ $m_{\rm F275W}-m_{\rm
        F336W}$ (left) and $m_{\rm F336W}$ vs.\ $m_{\rm F336W}-m_{\rm
        F435W}$ (right).  We have colored in green and magenta the two
        groups of stars selected in Fig.~\ref{allsel}.  This is the
        first time anyone has been able to follow two stellar
	populations
        in a globular cluster from the main sequence to the horizontal
        branch.} 
         \label{CMDpops}
   \end{figure*}

The behavior of the two groups is strikingly similar from the MS all the
way to the HB. Each of these stages shows a similar 80/20 ratio of star
numbers in the two groups, in the central field, and a similarly
decreasing radial trend of this ratio.  The most straightforward
interpretation is that we are seeing two populations that wind their
way, along nearly parallel paths, through the various successive stages
of stellar evolution. This continuity is pictured in Figure \ref{CMDpops},
whose two CMDs emphasize the crucial role that the F275W and F336W
filters play in seeing such simplicity in what would otherwise have been
a perplexing m\'elange of details.

Finally, the main properties of the two populations are summarized in
Table~\ref{popolazioni}.

\begin{table*}[ht!]
\begin{center}  
\caption{Chemical composition and fraction of stars relative to
  the total number, for the two main population groups.}
\scriptsize {
\begin{tabular}{cccccc}
\hline
\hline
Group & color code & sequences   & chemical composition &
fraction & fraction\\
      &            &             &                      &
R$<$$\sim$2 arcmin& R$>$$\sim$15arcmin\\
\hline
 {\it a}   & green   & MSa+SGBa+RGBa+HBa & CN-weak, O-rich, Na-poor, $Y$$\sim$0.25 & $\sim$20\% &$\sim$40\%  \\
 {\it b}  & magenta & MSb+SGBb+RGBb+HBb & CN-strong, O-poor, Na-rich, $Y$$\sim$0.265 & $\sim$80\% & $\sim$60\%\\
\hline
\hline
\label{popolazioni}
\end{tabular}
}
\end{center}
\end{table*}
%
%

\section{Summary and Discussion}
\label{sec:discussion}

We have analyzed a large set of {\it HST} and ground-based images of the
Galactic globular cluster NGC 104 (47 Tuc) in nine photometric bands,
finding multiple sequences throughout the various CMDs, from the main
sequence all the way to the horizontal branch.  Exploiting this wealth
of {\it HST} data to investigate the behavior of the multiple
populations as seen in several different combinations of magnitudes and
colors, we found that among the rainbow of possible CMDs, those
involving the F275W and F336W filters are particularly effective in
separating components of otherwise entangled cluster populations.  
Taking a cue from {\it HST}, we were able to construct a color system
based on the $U$ band that exhibited a similarly effective separation of
the populations in ground-based data.

We found that the distribution of stars along the MS, SGB, RGB, and HB
was in every case bimodal in the $m_{\rm F275W}-{\it m}_{\rm F336W}$
vs.\ $m_{\rm F336W}-m_{\rm F435W}$ two-color diagram, and we finally
put together the groups of stars that we had separated in this way, so
as to draw a continuous connection between their successive evolutionary
phases.

Near the cluster center all evolutionary phases 
split into two near-parallel sequences, with the richer one making up
about 80\% of the cluster stars, and the poorer one the remaining 20\%.
Wide-field ground-based photometry allowed us to identify and separate
the two sub-populations at larger radial distances from the cluster
center, with the result that the majority population is more centrally
concentrated, but its relative fraction decreases outward and approaches
50/50 in the outskirts of the cluster.  Globally, the majority
population accounts for $\sim 70\%$ of the whole population of 47 Tuc,
most of the remainder consisting of the minority population.  Radial
gradients in the stellar populations of this cluster have been known for
a long time, with
a CN-strong population more centrally concentrated than the CN-weak
one.  This suggests that the numerically dominant population is CN-strong.

Along these same lines, we used CN band strengths and Na and O
abundances that are available from the literature for some RGB and HB
stars of both populations, to investigate their chemical content.  It
appears that the more populous RGBb and HBb sequences consist of
CN-strong/Na-rich/O-poor stars, while the bulk of the
CN-weak/Na-poor/O-rich stars belong to the numerically poorer RGBa and
HBa.

On the theoretical side, we calculated synthetic spectra of
main-sequence stars with different chemical compositions, derived the
corresponding colors for our filter set, and compared them with the
observed colors in the two distinct populations.  The colors of the
minority population are well reproduced by stars with primordial helium
abundance and an oxygen-rich/nitrogen-poor composition that is typical
of halo stars of metallicity similar to that of 47 Tuc.  On the other
hand, the colors of the majority population stars are well reproduced by
a composition in which nitrogen is greatly enhanced, along with a
slightly increased helium, while carbon and oxygen are depleted.
Synthetic spectra for RGB and HB stars confirm this result.

The most straightforward interpretation of these differences is that the
minority population is the remnant of the first stellar  generation, 
which formed out of the interstellar medium of its
time, and shared its chemical composition.  The chemical composition of
the majority population, by contrast, carries the signatures of CNO and
proton-capture processing at high temperatures, such as depletion of
oxygen in favor of nitrogen and sodium, accompanied by helium
enhancement.  Therefore the majority population should be regarded as
the second stellar generation of 47 Tuc, which formed out of
material that had been partly processed through stars of the first
generation.

Since both populations share the same iron abundance, one can exclude
from the enrichment history of the second-generation material any
significant contribution by massive stars, exploding as core-collapse
supernovae.  This leaves intermediate-mass stars of the first generation
as the obvious candidates for having processed the material that is now
incorporated into the second generation, a view that is indeed widely
entertained in the literature.  Furthermore, it is certainly striking
that the second generation is today more populous than the first.
This sets strong constraints on the nature of the progenitor of
the present cluster 47 Tuc, and on the initial mass of the first
generation, which must have been one to two orders of magnitude more
massive than the portion that is still bound to the cluster.
 
Finally, we must again remind the reader that besides the two
populations that we have called first and second, we have found
unmistakable evidence for the presence of a third population, including
some $\sim 8\%$ of the stars, which 
we can distinguish it only on the SGB, but it appears to have the
abundance distribution of the ``b''  population.
This remains as a strident reminder that the stellar populations in 47
Tuc are more complex than the two-generation picture that we have just
sketched.

\begin{acknowledgements}
A.P.M., G.P., A.B., R.G., S.C., A.R., A.B., E.C., F.D., M.D.C., S.L.,
and A.P.\ acknowledge partial support by PRIN MIUR 20075TP5K9 and PRIN
INAF "Formation and Early Evolution of Massive Star Clusters. G.P.,
S.C., and 
A.R. acknowledge partial support by ASI under the program ASI-INAF
I/016/07/0.  J.\ A.\ and I.\ R.\ K.\ acknowledge support from STScI grant
GO-12311. 
\end{acknowledgements}
\bibliographystyle{aa}

\end{document}